\documentstyle[12pt,epsfig]{article} 
\def \LSP{\widetilde{N}_1}
\def \N2{\widetilde{N}_2}
\def \CH{\widetilde{\chi}^{\pm}}
\def \W1{\widetilde{\chi}_1^{\pm}}
\def \WC2{\widetilde{\chi}_2^{\pm}}
\def \SNU{\tilde{\nu}}
\def \BARSNU{\tilde{\bar{\nu}}}

\def \MLSP{M_{\widetilde{N}_1}}
\def \MN2{M_{\widetilde{N}_2}}
\def \MCH1{M_{\widetilde{\chi}_1^{\pm}}}
\def \M2CH{M_{\widetilde{\chi}_2^{\pm}}}
\def \MSNU{m_{\tilde{\nu}}}
\def \MSELL{m_{\tilde{e}_L}}
\def \MSELR{m_{\tilde{e}_R}}
\def \GLUM{M_{\tilde{g}}}
\def \MSQ{m_{\tilde{q}}}

\begin{document}
\setcounter{page}{0}
\thispagestyle{empty}

\begin{flushright}
TIFR/TH/96-27 \\
hep-ph/9605432
\end{flushright}

\begin{center}
{\LARGE\bf  VIRTUAL LSPs AT e$^+$ e$^-$ COLLIDERS\\}
\bigskip
{\normalsize Amitava Datta and Aseshkrishna Datta}\\
{\footnotesize 
Department of Physics, Jadavpur University, Calcutta 700 032,
India.}\\

{\normalsize Sreerup Raychaudhuri \\}
{\footnotesize 
Theoretical Physics Group, Tata Institute of Fundamental
Research, \\ Homi Bhabha Road, Mumbai 400 005, India.} \\
{\footnotesize\it Electronic address: sreerup@theory.tifr.res.in
\\}
\end{center}
\vskip 5pt

\begin{center}
{\large\bf ABSTRACT} 
\end{center}
\footnotesize
Currently popular search strategies for supersymmetric particles
may be significantly affected due to relatively light sneutrinos
which decay dominantly into invisible channels. In certain cases
the second lightest neutralino may also decay invisibly leading
to two extra carriers of missing energy (in addition to the
lightest supersymmetric particle (LSP) ) -- the virtual LSPs
(VLSPs). It is shown that if the sneutrino masses happen to be
in the small but experimentally allowed range ($\MSNU \approx$
45-55 GeV), these particles together with neutralino pairs may
contribute significantly to the missing energy in the process
$e^+ e^- \longrightarrow \gamma + \not \!\!{E_T}$ at {\rm LEP-2}
energies as an enhancement over the Standard Model or the
conventional MSSM predictions.  It is further shown that a much
larger region of the parameter space can be scanned at a high
luminosity $e^+ e^-$ collider at 500 GeV like the proposed NLC
machine.  Moreover this process can play a complementary role to
direct chargino searches at LEP-2 and NLC which may fail due to
a near mass degeneracy of the chargino and the sneutrino.
Formulae for the cross sections taking into account full mixings
of the charginos and the neutralinos are derived.  The signal
remains observable even in the context of more restricted models
based on  $N$=1 SUGRA  with common scalar and gaugino masses.
The effect of soft photon brehmsstrahlung on the signal is also
discussed briefly.
\normalsize\rm

\newpage
\section{Introduction}

It is well known that supersymmetry (SUSY) \cite{R1} is an
attractive alternative to the Standard Model (SM) since it
offers an elegant solution of the notorious naturalness problem,
provided the masses of the superpartners are of the order of 1
TeV or less. The search for SUSY at the TeV scale is, therefore,
a high-priority programme of current high energy physics.
Extensive searches for SUSY at the present high energy
accelerators including the Fermilab Tevatron and LEP-1  and
LEP-1.5 have yielded negative results and have eliminated
certain regions of the parameter space of  the Minimal
Supersymmetric extension of the Standard Model (MSSM)\cite{R1}.

However, there are small but interesting regions of the
parameter space, which are allowed by all experimental data,
where the signatures of SUSY can be significantly different from
the conventional ones considered in most cases. As an example,
let us note that in most cases the search strategies for
$R$-parity conserving  SUSY particles are based on the
assumption that there is a single, stable, weakly-interacting
neutral superparticle, the so-called lightest supersymmetric
particle(LSP). This particle, if produced, easily escapes
detection and carries missing transverse energy ($\not \!\!
E_T$). Moreover, as a result of $R$-parity conservation, all
other superparticles decay into the LSP either directly or
through cascades. Thus any sparticle production is accompanied
by $\not \!\!  E_T$, traditionally regarded as the most powerful
weapon in the arsenal of  SUSY hunters, carried by the LSP
alone.

It has been emphasised in recent literature \cite{R2,R3} that in
some interesting regions of the parameter space of the MSSM
(with $R$-parity conservation) there could be other carriers of
missing energy in addition to the LSP, due to sparticles which
decay dominantly into invisible modes. In such a scenario the
signals of sparticle production can be considerably different
from the conventional ones. This happens in the following
scenario.

The MSSM contains four spin-$\frac{1}{2}$ neutral particles.
These particles are the superpartners of the photon, the
$Z$-boson and the two neutral $CP$-even Higgs bosons. Linear
combinations of these four states, the four neutral gauginos or
neutralinos ($\widetilde{N_i}$, i=1,4), are the physical states.
In the currently favoured models, the lightest
neutralino($\LSP$) is assumed to be the LSP \cite{R1}.
Similarly, linear combinations of the superpartners of the
$W$-boson and the charged Higgs boson give two physical charged
gauginos or charginos.

The usual assumption that the MSSM is embedded into some Grand
Unified Theory (GUT) immediately implies, irrespective of the
choice of any particular gauge group for the GUT, that the
masses and the couplings of charginos and neutralinos depend
only on three independent parameters. Usually these are taken as
$\mu, \tan \beta$ and the gluino mass $\GLUM$.

If no further assumption is made then the masses of the
sfermions are totally independent of the gaugino-masses (we
shall discuss below more restricted  models with additional
theoretical assumptions).  Thus the sneutrinos ($\SNU$, the
superpartners of the neutrinos), though heavier than the LSP,
could very well be  lighter than the lighter chargino ($\W1$),
the second lightest neutralino ($\N2$) and other superparticles.
As a consequence, the invisible two-body decay mode $\SNU
\longrightarrow \nu \LSP$ opens up and completely dominates over
the others, being the only kinematically-allowed two-body decay
channel for the sneutrinos. The other necessary condition for
this scheme to work is that the $\LSP$ has a substantial zino
(superpartner of the $Z$-boson) component.  This, however, is
almost always the case as long as the gluino ($\tilde{g}$, the
superpartner of the gluon) has a mass ($\GLUM$) in the range
interesting for the SUSY searches at the Tevatron \cite{R4}.
Moreover, in such cases the $\N2$ --- which also has a dominant
zino component --- decays primarily through the process $\N2
\longrightarrow \nu \SNU$. This, however, also requires the left and
the right handed sleptons($\tilde{l_L}$ and $\tilde{l_R}$, the
superpartners of leptons) to be heavier than $\N2$. These two
particles ($\N2$ and $\SNU$), decaying primarily into invisible
channels, may act as additional sources of $\not \!\! E_T$ and
can significantly affect the strategies for SUSY searches
\cite{R2,R3}.  They are, therefore, called {\it virtual} or {\it
effective} LSPs (VLSPs or ELSPs) \cite{R2,R7} in the subsequent
discussion.

Some consequences of the VLSP scenario (as opposed to the
conventional MSSM where the LSP is the only source of missing
$E_T$) in the context of SUSY search at both hadron and $e^+ \,
e^-$ colliders  have been discussed in the literature
\cite{R2,R3,R5,R6,R7}. Here we wish to reiterate that for LEP
experiments beyond the $Z$-pole the predictions of the VLSP
scenario are significantly different from the conventional ones.
For example, experiments at LEP-1.5 \cite{R8} have recently
reported some improved bounds on the chargino-neutralino sector.
These bounds are derived from the processes ($a$) $e^+e^- \,
\longrightarrow \, \LSP \N2$ and ($b$) $e^+e^- \,
\longrightarrow \, \widetilde{\chi}^+_1 \widetilde{\chi}^-_1$,
assuming that $\widetilde{\chi}^{\pm}_1$ and $\N2$ primarily
decay into 3-body channels as predicted by the MSSM.  In the
VLSP scenario, however, the final state of process ($a$) is
invisible. Thus the improved bounds on the neutralino sector
from LEP-1.5 are not applicable in this scenario. Similarly in
the presence of light $\SNU$-s, $\W1$ primarily decays (with
branching ratio $\simeq$1) into the hadronically quiet channel
$l^{\pm} \SNU$ \cite{R2,R5,R7}. Thus the bounds on the chargino
sector derived from the absence of events containing acoplanar
jets and leptons and missing energy may have to be revised in
this scenario. It will be interesting to use the absence of two
acoplanar leptons in the above experiments to constrain the
($\MCH1 - \MSNU$) mass plane in the VLSP scenario.  However the
constraints thus obtained will depend on $\MCH1 - \MSNU$ and
can be completely evaded if $\SNU$ and $\W1$ are nearly
degenerate so that the leptons in the final state are soft and
unobservable.  We now focus on a signal in the VLSP scenario
which may lead to viable signals or constraints inspite  of such
a small mass  splitting.

We consider the process $e^+e^- \, \longrightarrow \, \gamma \,
+ $nothing($\not \! \! {E_T}$), already discussed in a previous
letter \cite{R6} in the context of {\bf LEP-2}. Here we shall
discuss the signal both at LEP-2 and at other future  $e^+ e^-$
colliders at high energies.  In the SM only  $\nu
\overline{\nu}$ pairs contribute to the final state.  In the
conventional MSSM both $\nu \overline{\nu}$ and $\LSP \LSP$
pairs contribute to this kind of effect.  With VLSPs, however,
there will be additional contributions from $\SNU \BARSNU$ and
$\widetilde{N_i}\widetilde{N_j}~(i, j = 1,2)$ which tend to
increase the cross section quite significantly.  In  \cite{R6}
it was found that a significant enhancement of the cross section
over the prediction of the SM occurs at LEP-2 in a reasonable
region of the MSSM parameter space (see section 2 for the
details)  allowed by the experimental data (most notably from
LEP-1 \cite{R9}).  Moreover, the bulk of the extra contribution
comes from $\SNU\BARSNU$ pairs.  Thus, such a signal, if
detected, can be distinguished not only from the SM but also
from the conventional MSSM without VLSPs.

In this work we have elaborated the results of \cite{R6} with
further details. The scan over the LEP-1 allowed parameter space
is now more complete. This, however, does not alter the results
of \cite{R6} qualitatively, although some quantitative changes
are noted. Assuming a conservative detector design as in
\cite{R6} we have found that at LEP-2 the statistical
significance of the signal is rather modest.  For optimistic
choices of SUSY parameters(most notably for relatively low
sneutrino and gluino masses, $\MSNU$= 45---60 GeV, $\GLUM
\simeq$ 200GeV), signals with statistical significance $ \geq 3
\sigma$ can be obtained (numerical details are given in the next
section).

The cross sections for the process $e^+e^- \longrightarrow
\gamma + nothing (\not \!\! E_T)$ has been discussed extensively
in the literature.  We have done a complete calculation in the
VLSP scenario without using the simplifying assumptions used in
earlier works. In \cite{R6} we presented some of the numerical
results. But the formulae for the cross sections, which are
quite cumbersome, could not be presented in a brief letter. A
major result of this paper is the detailed formulae presented in
a compact form. First  we have calculated the full cross section
for the purely SM process $e^+ e^- \longrightarrow \gamma \nu
\bar{\nu}$. In many of the earlier works \cite{R10}, appropriate
for LEP-1, the contribution of the $W$-exchange diagrams was
computed in the limit of four-fermion contact interaction. We
have recalculated it with the full $W$-propagator. We have also
taken  the widths of $W$ and $Z$ into account. Our results agree
completely with those of \cite{R10} after taking the appropriate
limits. In \cite{R11} this cross section was also computed
without any approximation.  However, the published results
include several misprints (see, for example, equation(3) which
contains several terms which are dimensionally incorrect).  This
makes comparison rather difficult. This cross section was also
computed in \cite{R12} by neglecting the widths but keeping the
full $W$-propagator. Their analytical formulae agree completely
with ours in the appropriate limit. Moreover, a comparison of
the numerical results shows that effects of the widths are
indeed negligible, at least for the energy ranges considered in
this paper.

The most important contribution to this process in the VLSP
scenario comes from  $e^+ e^- \longrightarrow \gamma \SNU
\BARSNU$. Only the amplitudes of the relevant Feynman diagrams
are given in, for example, \cite{R13} in the limit when the
chargino is purely a wino (superpartner of the $W$-boson).  We
have computed the full cross section taking into account the
chargino-mixing matrix. Our numerical results agree with those
of \cite{R13} in the appropriate limit.  In a more recent paper
\cite{R18} this cross section has been computed by assuming the
charginos to be very massive. In this limit we  agree with the
main features of their results.

We have also computed the cross section for the process $e^+ e^-
\longrightarrow \gamma \widetilde{N_i} \widetilde{N_j}$, $(i,j = 
1,2)$ taking the $4\times 4$ neutralino mass matrix into
account. This cross section with only LSP-pairs ($i = j = 1$) in
the final state is also relevant for the conventional MSSM and
was computed in \cite{R14} in the limit when the  $\LSP$ is a
pure photino without any mixing. In this approximation the
s-channel $Z$-exchange diagrams are absent which reduces the
number of diagrams and  interferences between them. Our
numerical results  agree, in the appropriate limit, with the
those of \cite{R14}. In a very recent paper \cite{R15} the
calculation for a mixed LSP has been done using the structure
function approach \cite{R16}. One of the conclusions of
\cite{R6}, {\it viz.} LSP pairs alone cannot give a signal with
acceptable statistical significance, is supported by \cite{R15}.
The general formulae presented in this paper also include the
contribution of $\widetilde{N_1} \widetilde{N_2}$ and
$\widetilde{N_2} \widetilde{N_2}$ pairs.

In Appendix-A we  present analytical formulae for all the matrix
elements squared.

Using these results we have also computed the cross sections at
$e^+ \, e^-$ colliders at higher energies after introducing
kinematical cuts to reduce the SM backgrounds. Many of these
machines are likely to be of very high luminosities\cite{R17}.
As a consequence of this, signals of very high statistical
significance ( $>5 \sigma$) can be obtained at CM energies
$\approx$350 and 500 GeV which are attainable at the proposed
Next Linear Collider(NLC).  Special care, however, should  be
taken to reduce the background from  radiative Bhabha scattering
where both the final state charged particles are lost in the
beam pipe \cite{R18}. This will be discussed in further details
in section 3.

The VLSP scenario, which is certainly consistent with all
available experimental results on SUSY searches, can also be
accommodated in the more constrained and theoretically motivated
models based on $N=1$ Supergravity with common scalar and
gaugino masses at a high scale \cite{R19}. In this scenario the
sneutrino  and the gaugino masses are not completely
independent, but get related through the renormalisation group
(RG) equations. It was shown in \cite{R7} that the VLSP scenario
can be accommodated even in this highly constrained scenario. In
this paper we have found that the signal at LEP-2 is reasonable
for certain regions of the  parameter space  given in \cite{R7},
while at NLC, signals with high statistical significance can
still be obtained..

It may be noted at this point that in \cite{R6} only the lowest
order cross section was considered. In this paper we have
considered the effects of soft photon brehmstrahlung on the
cross section. These effects can be obtained to all orders in
perturbation theory by using, e.g., the structure function
approach of \cite{R16}. We have found that the impact of these
corrections on the signal is  rather modest for the entire
energy range considered by us and it leaves the signal to
(root)background ratio almost unaffected.

The plan of the paper is as follows. In section 2 we consider
the signal at $\sqrt{s}$=190 GeV corresponding to the LEP-2
energies and briefly comment on the possibilities at LEP-1.5  In
section 3 the same discussion is carried out for high luminosity
$e^+e^-$ colliders operating at higher energies. In section 4
the signal is discussed in the context of highly constrained
models based on $N=1$ SUGRA.  Section 5 discusses briefly the
effect of soft photon brehmsstrahlung on the signal.  Our
conclusions are summarised in section 6. The relevant formulae
for the cross sections are given in the Appendix.

\section{The Signal at LEP-2 Energies }

In our calculations we use the usual assumption of a common
gaugino mass at the GUT scale which relates the U(1) gaugino
mass $M_1$ with the SU(2) gaugino mass $M_2$ \cite{R1}. With
this assumption $M_2$, the higgsino mass parameter $\mu$ and tan
$\beta$ determines the masses and the couplings of the charginos
and neutralinos completely. In some of our figures we have used
$3 M_2$ as the variable instead of $M_2$. The advantage is that
this is approximately equal to the running gluino mass (through
the above assumption of unification)  which can be directly
related to the constraints obtained from the Tevatron \cite{R4}.
We, however, emphasise that this equality is only approximate
and the factor 3 may change, though not drastically, with the
energy scale. Nevertheless we shall denote in the following $3
M_2$ by the gluino mass ( $\GLUM$ ) for the sake of simplicity.

In addition we have assumed the  $SU(2)$ breaking relation :
\[ \MSELL \, = \, \sqrt{\MSNU^2 \, + \, {\rm cos}^2 \theta_W \, D_Z}
\] where \[ D_Z \, = \, M_Z^2 \, \frac{{\rm tan}^2 \beta -1}{{\rm
tan}^2 \beta +1} \] and $\MSNU$ is treated as a free parameter
and three degenerate sneutrinos are assumed.  For the right
handed sleptons we have made the popular assumption $\MSELR
\approx \MSELL$ although  deviations from this approximation may
naturally occur in some models.

In the VLSP scenario the following constraints must be satisfied
\cite{R7}:
\begin {eqnarray*}
\MSNU < \MN2 < \MSELL, \MSELR \\
\MSNU < \MCH1 < \MSELL
\end{eqnarray*}

In Figs.[1A] and [1B] we present the regions in the
($\GLUM-\MSNU$) mass plane ( where the precise definition of the
parameter $\GLUM$ is given in the first paragraph) compatible
with the above inequalities for $\mu=-\,$250 (Fig.1A) or +250
(Fig.1B) GeV, tan$\beta$=10, 150 GeV $\leq \GLUM \leq$ 800 GeV
and $\MSELL=\MSELR$. In each figure the entire bounded area
corresponds to the region of the parameter space where
$\tilde{\nu}$-s behave like  VLSP-s.  Corresponding to each
$\MSNU$, this happens for a range of $\GLUM$. The lower limit of
this range comes from the condition $\MSNU<\MN2,\MCH1$ while the
upper limit comes from $\MLSP<\MSNU$. If the additional
condition $\MN2<\MSELL,\MSELR$ is satisfied then $\N2$ also
decays invisibly.  This happens in the shaded region of the
figures. The area of this region, however, is crucially
dependent on the choice $\MSELL \approx \MSELR$. If $\MSELR$ is
reduced, the shaded areas may shrink further. From Figs.[1A] and
[1B] it is also apparent that the allowed region is almost
independent of the choice of $\mu$.

As discussed in the introduction, the processes ({\bf A})
$e^+e^- \longrightarrow  \widetilde{\nu} \widetilde{\bar{\nu}}
\gamma$, ( {\bf B}) $e^+e^-  \longrightarrow  \LSP \LSP \gamma$ and
({\bf C}) $e^+e^-  \longrightarrow  \LSP \N2 \gamma$ contribute
to the signal at $\sqrt{s}=190$ GeV.  At this energy the
contribution of $e^+e^-  \longrightarrow  \N2 \N2 \gamma$ is
indeed negligible. We have scanned over the entire LEP-1 allowed
parameter space compatible with the VLSP scenario and have
computed the cross sections from the processes {\bf A} and {\bf
B}. If, for a particular choice of the parameters, $\N2$ is also
a VLSP (the shaded region in Figs.[1A] and [1B]) then the
contribution from the process {\bf C} is also taken into
account.

We show in Fig.[2A] the energy distribution of the photon for
processes {\bf A}(the solid line) and {\bf B}(the dashed line).
Photons in the forward direction (those emitted within an angle
of 5$^\circ$ with the beam axis) are not considered. There is a
mild cut $E_{\gamma}>5$ GeV which in conjunction with the strong
angular cuts(discussed below) removes other backgrounds
including the ones from radiative Bhabha scattering with the
final state $e^+e^-$ pair going down the beam pipe. This cut
also takes into account the detector threshold. The SUSY
parameters used are $\MSNU = 50$ GeV, $\MSELL = \MSELR$,
$\GLUM=$ 200 GeV, $\mu=-$200 GeV and tan$\beta$= 5.

The main SM background comes from the process {\bf (D)}$ e^+e^-
\longrightarrow  \nu \bar{\nu} \gamma$. The corresponding distribution
(the dotted line) for this background is also shown in Fig.[2A].
As has already been discussed in \cite{R6}, this distribution
has a peak at about $(s-M_Z^2)/2 \sqrt{s}$ corresponding to the
decay of a real $Z$ into $\nu \bar{\nu}$, as expected. Thus an
upper cut of $E_{\gamma}<$60 GeV optimises  $\sigma =
\frac{S}{\sqrt{B}}$ where S is the number of signal events and B
is the number of background events.

In Fig.[2B] we present the angular distribution of the signal
(for the above SUSY parameters) and the background following the
conventions of Fig.[2A]. The distributions have similar
characteristics. Thus angular cuts cannot further improve the
quality of the signal. An irreducible background therefore
remains.

In order to make a conservative assessment of the prospect of
discovering the signal we impose an angular cut
$40^\circ<\theta_{\gamma}<140^\circ$, where $\theta_{\gamma}$ is
the angle between the photon and the direction of the positron.
This cut corresponds to the high $p_T$ photons collected in the
central part of the detector where photon detection efficiency
is expected to be large ($\approx 1$). We have also studied the
effects of a cut allowing for more angular coverage alongwith an
explicit strong cut on the $p_T$ of the photon to remove the
Bhabha background. These cuts, introduced in a recent paper
\cite{R15}, are given by $18^\circ<\theta_{\gamma}<162^\circ$,
$1<E_{\gamma}<47.5$ GeV and $p_{T_{\gamma}}> 6.2$ GeV. Now the
photons detected in the endcap region of the detector also
contribute and it is assumed that their detection efficiency is
still large($\approx 1$). We compare the response of the signal
for the two sets of cuts in Table-I and find that they give very
similar results. Using our cuts the background is 0.45 pb while
for the cuts of \cite{R15} it is 0.51 pb.

In Fig.[3] we present the cross section as a function of
$m_{\tilde{\nu}}$ for $\MSELL = \MSELR$ with our conservative
cuts.  The other SUSY parameters are varied within the following
range: 200$\leq \GLUM \leq$400, $-$500$\leq \mu \leq$500, 2$\leq
tan \beta \leq$30. Only points allowed by LEP-1 data are
considered. This scanning of the parameter space is more
comprehensive than the one carried out in \cite{R6}.  We note
that $\GLUM < 212$ GeV is ruled out from SUSY searches at
Tevatron for $\MSQ \approx \GLUM$ \cite{R4}.  For $\MSQ >>
\GLUM$, the limit is $\GLUM >$144 GeV. Strictly speaking these
limits pertain to the pole mass of the gluino. However for $\MSQ
\approx \GLUM$, the running gluino mass ( the parameter more
directly related to the chargino - neutralino sector through the
assumption of unification ) is equal to the pole mass to a very
good approximation \cite{R22}. For $\MSQ >> \GLUM$, the
difference between the above two masses is significant. In fact
it turns out that in this case the above limit on $\GLUM$
translates into a much relaxed bound on the running gluino mass.

It has also  been qualitatively argued that these limits may be
relaxed in the VLSP scenario\cite{R2}. However, no quantitative
result exists. Moreover for low gluino masses the lighter
chargino masses are also reduced and it becomes increasingly
difficult to accommodate the VLSP scenario. We have, therefore,
not assumed any drastic reduction of these mass limits and have
taken conservatively $\GLUM \geq$200 GeV.  As discussed above
our $\GLUM$ roughly corresponds to the running mass and is
defined precisely at the beginning of this section.  However the
squark mass is essentially a free parameter in this model
independent analysis and does not affect directly any of our
numerical results. Sometimes therefore we shall take mainly for
the purpose of illustration $\GLUM$ around 150 GeV which is
allowed for heavy squarks. For $\GLUM \geq$400 GeV the signal
falls below the 2$\sigma$ level and becomes uninteresting. The
band within the solid lines in Fig.[3] corresponds to the
combined cross sections $\sigma_{tot}\,$ from the processes {\bf
A}, {\bf B} and {\bf D}, i.e. the scenario in which the $\SNU$
is the only VLSP. In order to obtain conservative estimates, we
have not considered the possibility that $\N2$ may also be a
VLSP. This is because the latter possibility can be evaded by an
appropriate choice of $\MSELR$.  The width of this band is due
to varying $\GLUM$, $\mu$ and tan$\beta$ within the above ranges
and is a measure of SUSY parameter space consistent with the
VLSP sceanrio for a given $\MSNU$. Taking into account the
points where $\N2$ is also a VLSP, for the choice
$\MSELL=\MSELR$, the signal improves modestly due to the
contribution from process {\bf C} which is shown by the band
enclosed by the dashed lines.  For comparison we also display
the cross section for the background (process {\bf D}) which
corresponds to the lowest dotted horizontal line in the figure.
The other two dotted horizontal lines correspond to 3$\sigma$
and 5$\sigma$ fluctuations of the background events for an
integrated luminosity of 500 pb$^{-1}$. From the figure it may
be noted that a 5$\sigma$ signal can be obtained for $\MSNU
\leq$52 GeV for a very small region of SUSY parameter space. A
much larger region of parameter space gives events above
3$\sigma$ fluctuation with $\MSNU \leq$66 GeV. In this figure
the upper edges of the bands correspond to $\GLUM =$ 200 GeV and
low tan$\beta$ (2$\leq$tan$\beta \leq$6).  The cross section is
rather insensitive to the variation of $\mu$.  For $\GLUM =$ 300
GeV only 3$\sigma$ signals can be obtained for a limited region
of the parameter space while for $\GLUM =$ 400 GeV the signal
remains below the 2$\sigma$ level for the entire region of the
parameter space.

A clearer representation of the regions in the ($\GLUM-\MSNU$)
plane that can be probed at $\sqrt{s}$=190 GeV is given by the
contour plot in Fig.[4] for three values of tan$\beta$=2,10,30.
The points within the solid, dashed and dotted contours yield
signals with statistical significances $\geq$4$\sigma$,
3$\sigma$ and 2$\sigma$ respectively for suitable choices of
$\mu$.  In this figure we have also considered signals in the
region  150 GeV $ \leq \GLUM \leq $200 GeV, since this region is
still allowed by the Tevatron data for heavy squarks. The
signal,however, is not considered for points where the VLSP
constraints discussed above are not satisfied.

Since experiments at LEP-1.5 are in progress, the cross section
at $\sqrt{s}=$130 GeV is of considerable interest. However, even
for $\MSNU=$50 GeV and other favourable choices of the SUSY
parameters ($\mu=-$300, $\GLUM=$200 and tan$\beta=$5) the cross
section happens to be rather disappointing. The total cross
section of the processes {\bf A}$-${\bf C} is 0.054 pb while the
background is 0.373 pb with the cuts $5<E_{\gamma}<20$ GeV and
$40^\circ<\theta_{\gamma}<140^\circ$. Thus for an integrated
luminosity of 6 pb$^{-1}$ the statistical significance is
certainly $<3{\sigma}$.

From the above results it is clear that the process under
consideration has a rather modest cross section at LEP. The bulk
of the constraints in the chargino-sneutrino sector in the VLSP
scenario is therefore likely to come from direct chargino
searches \cite{R7}. Nevertheless this process is likely to play
a complementary role in  the regions of the parameter space
where the chargino and the sneutrino are nearly degenerate and
the sneutrino masses are relatively small.  To illustrate this
point we consider an example with $\mu$ = -350 GeV, tan $\beta$
= 6 and $\GLUM$ = 150 GeV. In this case the chargino mass is 53
GeV. The total signal cross section for $\MSNU$ = 50 GeV is 0.17
pb which corresponds to 5 $\sigma$. It can be readily checked
that the signal is weaker for higher $\GLUM$s, i.e., for larger
chargino - sneutrino mass differences. This reduction for higher
$\GLUM$ is essentially due to propagator supression  in process
A) and due to kinematical effects in processes B) and C) and
holds for other choices of SUSY parameters. Thus in contrast to
the direct chargino searches, the  regions of the parameter
space where the chargino and the sneutrino are nearly mass
degenerate can be probed via this mode, provided the sneutrinos
are not too heavy.  In Fig.4 the shaded region corresponds to
$\MSNU < \MCH1 < \MSNU + 5$ GeV, where the chargino decay is
likely to be difficult to detect. The statistical significance
of the signal cannot be judged directly from this figure for the
entire shaded region. This is because we have not computed the
cross sections for $\GLUM <$ 150 GeV, since this region is
either ruled out or marginally allowed by the Tevatron data
depending on the assumption on the squark mass.  The cross
sections have also not been computed for parameters for which
the VLSP condition is not satisfied. However in regions not
excluded by these considerations, the cross section is
significant. This is especially so for relatively large  $\tan
\beta$.

\section{The Signal at NLC}

In this section we discuss the signal and the background for
$e^+e^-$ collisions at {\bf NLC} for two values of centre of
mass energy {\it viz.} $\sqrt{s}$=350 GeV and $\sqrt{s}$=500
GeV.

For $\sqrt{s}$=350 GeV the energy and angular distributions of
the radiated photon in the signal [process {\bf (A)} and process
{\bf (B)}] and background [process {\bf (D)}] are shown in
Fig.[5A] and Fig.[5B] respectively. The set of SUSY parameters
used are $\MSNU$=80 GeV and $\GLUM$=350, $\mu=-$500 and
tan$\beta$=5. The conventions for different curves are the same
as those in Figs.[2A] and [2B]. The energy distribution
(Fig.[5A]) of the background has a peak at about $\sqrt{s}/2$
which is the beam energy. Thus an upper cut of $E_{\gamma}<$150
GeV is set.

A lower cut on the photon energy is set from naive kinematics of
the background due to radiative Bhabha scattering which requires
special care \cite{R18}. We have devised our cuts against this
background assuming that $e^+e^-$ scattered within a cone of
10$^\circ$ with respect to the beam axis may remain undetected.
From kinematics we find that in this situation the photons are
restricted by the following criteria: $E_{\gamma} \leq$ 65 GeV,
$p_{T_{\gamma}} \leq$ 52 GeV. We have also checked that either
of the above cuts reduces the Bhabha background completely. From
the $E_{\gamma}$ distribution (Fig.[5A]) it is also clear that a
strong lower cut of $E_{\gamma}>$ 65 GeV reduces both the signal
and the $\nu \bar{\nu}$ background from the process {\bf (D)},
but does not affect the $\sigma=\frac{S}{\sqrt{B}}$ ratio
drastically. The anticipated high luminosity (${\cal L}\sim
10^{33} cm^{-2} sec^{-1} \sim 3 \times 10^4 pb^{-1}$ over a year
) \cite{R17} ensures that a respectable number of events is
obtained in spite of this reduction due to stiff cuts.

Another set of cuts, subjected to a rather strong assumption
about the detectors, was discussed in \cite{R18}. In particular
it was assumed that it is possible to detect in a radiative
Bhabha event the scattered $e^+$ or $e^-$ emitted at an angle
$\theta_{min}<\theta_e<$10$^\circ$ ($\theta_{min} \approx
1.6^\circ$ at $\sqrt{s}$=350 GeV). If this indeed is the case
then  the above stringent lower $p_T$ cut on the photon can be
significantly relaxed.  We shall compare below the effects of
these two sets of cuts. It turns out that with the milder  cuts
of \cite{R18} a larger region of the parameter space can be
probed.

From Fig.[5B] it is clear that the angular distributions for the
signal and background processes have similar characteristics.
Thus, as in section 2,  angular cuts cannot improve the quality
of the signal. Nevertheless we impose  conservatively angular
cuts of 40 $^\circ<\theta_{\gamma}<140^\circ$ which correspond
to the central region of the detector where photon detection
efficiency is expected to be very high ($\approx$1).

In Fig.[6] we present the cross section as a function of
$\MSNU$. The conventions are the same as those in Fig.[3] for
the bands and the horizontal lines. The  dashed band contains
the additional contributions from $\LSP \N2$ and $\N2 \N2$ pairs
at points where $\N2$ is also a VLSP.  As emphasised in section
2 , the latter contributions will be absent if
$m_{\tilde{e}_{L,R}}< \N2$. As $\MSNU$ increases the minimum
$\GLUM$ which can accommodate the VLSP scenario also increases
(see Fig.[1]). For example, at $\MSNU$=100 GeV, only $\GLUM
\geq$375 GeV are consistent with the VLSP scenario.  In addition
to the obvious kinematical effects, suppressions due to $\CH$
and $\tilde{e}_{L,R}$ propagators, therefore, tend to decrease
the signal with increasing $\MSNU$. Also, for larger $\GLUM$,
the contributions from $\widetilde{N}_i \widetilde{N}_j$ pairs
decrease due to kinematic effects. The reduction of the total
VLSP cross section with increasing $\MSNU$ is therefore a
complicated combination of several effects. It is clearly seen
from Fig.[6] that the SUSY parameter space consistent with the
VLSP scenario, indicated by the width of the bands, gradually
shrinks as $\MSNU$ increases.

We find that for $\MSNU \leq$110 GeV a 5$\sigma$ signal can be
obtained even with  our conservative cuts without imposing any
special requirement on the detectors. This unfortunately is much
smaller than the kinematic limit at $\sqrt{s}$=350 GeV. It is
therefore worthwhile to study the effects of the relaxed cuts
proposed in \cite{R18}. We compare the efficiencies of the two
sets of cuts in Table-II. The complex interplay between the
$\MSNU$ and $\GLUM$ in the VLSP scenario, discussed in the last
paragraph, is also clearly exhibited in Table-II which is drawn
for $\GLUM$=400 GeV. For this $\GLUM$, the $\N2$ is not a VLSP
for $\MSNU$= 110 and 125 GeV, which leads to sleptons lighter
than the $\N2$. The cross section is, therefore, larger for
heavier sneutrinos at this $\GLUM$.  It follows from this table
that significantly larger regions of the parameter space can be
scanned if improvement in instrumentation discussed in
\cite{R18} allows the scattered $e^+e^-$ in a radiative Bhabha
event to be tracked down in the beam pipe.

In Figs.[7] we present contour plots in the ($\GLUM-\MSNU$)
plane that can be probed at $\sqrt{s}$=350 GeV for three values
of tan$\beta$, tan$\beta$=2,10,30. In these, the
dotted(outermost) contours represent the areas in the
($\GLUM-\MSNU$) plane where a $\geq$3$\sigma$ signal can be
obtained. The dashed(middle ones) and the solid(innermost ones)
show the same for 4$\sigma$ and 5$\sigma$ signals respectively.
As $\MSNU$ increases, a distinct rise in the lowest allowed
$\GLUM$ is also a very indicative feature of the VLSP scenario.

The photon energy and angular distributions for $\sqrt{s}$=500
GeV are shown in Fig.[8A] and Fig.[8B] respectively. The
conventions and features of the curves are similar to ones for
$\sqrt{s}$=350 GeV case. An upper cut of $E_{\gamma}<$225 GeV is
set. From kinematical considerations a strong lower cut of
$E_{\gamma}>95$ GeV is imposed to  eliminate completely the
radiative Bhabha background. Along with this an angular cut of
40$^\circ<\theta{\gamma}<$140$^\circ$ corresponding to the
central region of the detector is imposed.

In Fig.[9] we present the cross section as a function of
$\MSNU$. The conventions are the same as in Fig.[6]. We find
that only for $\MSNU \leq$125 GeV  5$\sigma$ signals can be
obtained using our conservative cuts and optimistic choices of
SUSY parameters.  It is again much smaller than the kinematic
limit at $\sqrt{s}$=500 GeV.  Once again by using the relaxed
cuts proposed in \cite{R18}, the search limit can be
significantly increased. We compare the efficiencies of the two
cuts in Table-III which shows the prospect of improvement in the
search limit if the relaxed cuts \cite{R18} are permissible due
to improvements in detector designs.

In Figs.[10] we present the contour plots at $\sqrt{s}$=500 GeV
for three values of tan$\beta$. The conventions are exactly the
same as in Fig.[7].

To end this section it is noteworthy that a greater region in
the ($\MSNU-\GLUM$) plane can be probed at $\sqrt{s}$=500 GeV
with appreciable statistical significance compared to the
$\sqrt{s}$=350 GeV case, as expected. However, this gain is not
commensurate with the increase in beam energy.  Also the
searches at NLC via this mode will be very effective in
constraining  the  regions of the parameter space which are  in
principle accessible to direct chargino searches at LEP 2
energies but can not be probed there due to near mass degeneracy
of the chargino and the sneutrino and relatively large mass of
the sneutrino.

\section{The Signal in $N$=1  SUGRA Models}

In this section we consider a more constrained scenario based on
$N$=1 SUGRA with a common scalar mass ($m_0$) at the GUT scale
\cite{R19}.  It should, however, be noted that recently many viable
models with non-universal scalar masses have been constructed
\cite{R20}. Yet models with a common $m_0$ continue to be popular and
its implication for the VLSP scenario is worth investigating.
However, no assumption about the Higgs sector and, consequently,
about the $SU(2)\otimes U(1)$ breaking mechanism is made. As
pointed out in \cite{R7} the VLSP scenario can also be
accommodated in this more restrictive model. It was shown that
the VLSP constraints require a relatively light gluino with
$\GLUM$  bounded by the relation \[ m_{1/2} \leq 1.4\sqrt{D_Z}
\] where $m_{1/2}$ is the common gaugino mass at the GUT scale
and $D_Z$ has been defined earlier. This bound also restricts
the masses of $\W1$ and $\N2$ severely. Since in the VLSP
scenario the sneutrino has to be lighter than the above
particles, $\MSNU$ is also bounded from above. As a consequence,
this scenario can be tested conclusively at relatively low
energy machines, e.g.,  at an $e^+ \, e^-$ collider
$\sqrt{s}$=350 GeV.  Sizable cross sections may be obtained at
$\sqrt{s}$=190 GeV provided $\MSNU$ happens to be  in the lower
part of its allowed range.

Here we consider the allowed region of the ($m_0-m_{1/2}$) mass
plane in the VLSP scenario as given in \cite{R7}. Using the
formulae in \cite{R7} one can calculate the sparticle masses and
hence the cross sections at various  points of the above region
using the cuts stated in the earlier sections. We present some
of the sample results at $\sqrt{s}$=190 GeV and $\sqrt{s}$=350
GeV in Table-IV. It is seen from Table-IV that the entire region
of the parameter space allowed in the VLSP scenario gives an
observable signal($\geq$ 5$\sigma$) at $\sqrt{s}$=350 GeV. In
each case, it turns out that $\MSQ$ is nearly equal to $\GLUM$.
Using the bounds of \cite{R4} we have restricted ourselves to
the cases with $\MSQ=\GLUM \approx$ 200 GeV.

If one further assumes radiative breaking of $SU(2) \otimes
U(1)$ symmetry then the number of free parameters reduces
further. In particular $\mu$ becomes a fixed parameter, apart
from a sign ambiguity, for given $m_0$, $m_{1/2}$, tan$\beta$
and $m_t$. We have already seen that the cross sections are not
very sensitive to $\mu$.  We therefore work with the
representative choice $\mu=-\GLUM$ used by other authors
\cite{R21}. The allowed regions of \cite{R7} now reduce to
narrow strips. In Table[IV] we give the cross sections at a few
representative points and note that observable signals with high
statistical significances are predicted at $\sqrt{s}$=350 GeV.

\section{Radiative Corrections}

In this section we briefly consider the radiative corrections to
the cross sections. We follow the structure function approach of
\cite{R16}. We, however, restrict ourselves to the corrections due to
soft photon brehmstrahlung only to all orders in perturbation
theory.  The formula for the corrected cross section can be
found in equation(18) of the fourth paper of \cite{R16}. In this
formula we have substituted the cross sections given in the
Appendix. At LEP-2 energies the background reduces from 0.45 pb
to 0.34 pb. The changes in the signal cross sections are shown
in Table-V. For each sneutrino mass the cross sections with and
without the radiative correction are presented using Cut{\bf A}
of Table-I. It is seen that both the signal and the background
reduce sizably due to this correction but the ratio $\sigma$
remains almost unaffected. It may also be noted that inspite of
this reduction the total number of signal events remains
adequate for a luminosity of 500 pb$^{-1}$.

\section{Conclusions}   

Following our earlier works \cite{R2,R5,R6,R7}, we have
emphasised in this paper  that currently popular search
strategies for supersymmetric particles  may be significantly
affected in the VLSP scenario with  relatively light sneutrinos
and the second lightest neutralino,  which may decay dominantly
into invisible channels, leading to two extra carriers of
missing energy (in addition to the lightest supersymmetric
particle (LSP)).

We have focussed our attention on the processes ({\bf a})
$e^+e^- \longrightarrow  \widetilde{\nu} \widetilde{\bar{\nu}}
\gamma$, ({\bf b}) $e^+e^-  \longrightarrow \widetilde{ N_i}
\,\widetilde{N_j} \gamma$ ($i,j$=1,2) and ({\bf c}) $ e^+e^-
\longrightarrow  \nu \bar{\nu} \gamma$, all of which contribute
to  $e^+ e^- \longrightarrow \gamma + \not \!\!{E}$ in the VLSP
scenario.  Formulae for the cross sections taking into account
full mixings of the charginos and the neutralinos are derived
and presented in the Appendix.

It is shown that for suitable choices of SUSY parameters at
LEP-2 energies, process (a) contributes dominantly to the
signal. The contribution from $\widetilde{N}_1
\,\widetilde{N}_2$ pairs is also comparable to that from LSP
pairs, which alone produces the signal in the conventional MSSM.
This leads to the interesting possibility that at LEP-2 the VLSP
scenario can be distinguished not only from the SM but also from
the conventional MSSM. If the sneutrino masses happen to be in a
small but experimentally interesting range $\MSNU = 45-55$ GeV
and the gluinos are relatively light ($\GLUM \approx$ 200 GeV),
sneutrino and neutralino pairs (processes (a) and (b)) may
contribute significantly to the signal as an enhancement over
the SM (process (c)) or the conventional MSSM (process (b) with
$i=j=$1 ) predictions. After suitable kinematical cuts
\cite{R6,R15}, the statistical significance of the signal
depends crucially on the choice of SUSY parameters and may be
$\geq 5 \sigma$ for $\MSNU$ and $\GLUM$ as above.

It is further shown that a much larger region of the parameter
space can be scanned at a high luminosity $e^+ e^-$ collider at
350 and 500 GeV like the proposed NLC machine even with
conservative cuts. If further improvements in detector design
\cite{R18} allow a relaxation of these strong cuts, even larger
regions of the parameter space can be scanned.  At
$\sqrt{s}$=350(500) GeV, $\MSNU \approx$ 150(180) GeV may be
probed for $\GLUM$ in the range 400--550 GeV using the cuts of
\cite{R18}. This signal is viable even if the mass splitting
between the chargino and the sneutrino is small and thus, can
play a complementary role to direct chargino searches at LEP-2
and NLC. The latter process can probe larger regions of the
parameter space, but the signal may disappear if the above mass
splitting is small.  The signal remains observable even in the
context of more restricted models based on  $N$=1 SUGRA  with
common scalar and gaugino masses at a high scale. The effect of
soft photon emmisions to all orders in perturbation theory is
also discussed briefly using the approach of
\cite{R16}. 
 
\subsection*{Acknowledgements:}

The authors wish to thank Prof. D.Roy, Jadavpur University for
help in computation.  They also wish to thank the organizers of
the Workshop on High Energy Physics Phenomenology (WHEPP4), held
in Calcutta in January 1996, where this work was discussed and
Manuel Drees for valuable comments.  Amitava Datta's work is
supported by the Department of Science and Technology,
Government of India, Project No. SP/S2-K-42/88.  The work of
Aseshkrishna Datta has been supported by the Council for
Scientific and Industrial Research, India. Sreerup Raychaudhuri
acknowledges the grant of a project (DO No. SR/SY/P-08/92) of
the Department of Science and Technology, Government of India.

\newpage

\def \MSNU{m_{\tilde{\nu}}}
\def \MSNUSQ{m_{\tilde{\nu}}^2}
\def \MZSQ{M_Z^2}
\def \CVSQ{C_V^2}
\def \CASQ{C_A^2}
\def \MASQ{m_a^2}
\def \MBSQ{m_b^2}
\def \CLEFT{(C_V-C_A)}
\def \VASQ{|{\cal V}_{a1}|^2}
\def \VBSQ{|{\cal V}_{b1}|^2}
\def \MZSQ{M_Z^2}
\def \CVSQ{C_V^2}
\def \CASQ{C_A^2}
\def \GASQ{G_A^2}
\def \CLEFT{(C_V-C_A)}
\def \MI2{m_i^2}
\def \MJ2{m_j^2}
\def \MSEL2{m_{{\tilde{e}_R}}^2}
\def \NI{\widetilde{N}_i}
\def \NJ{\widetilde{N}_j}
\vskip 8pt

\begin{flushleft}
{\LARGE \bf Appendix}
\end{flushleft}
\vskip 20pt
\noindent
In this appendix we systematically present the relevant formulae
for calculating the cross sections of different processes.
Throughout this paper we use the following Standard Model
Parameters :\\

            $ \alpha = 1/128.8, \;
              G_F   = 1.16637 \, \times 10^{-5}, \;
              M_Z   = 91.187 \, GeV, \;
              M_W   = 80.22  \, GeV, \;
              {\Gamma}_Z   =  2.498 \, GeV, \;
             {\Gamma}_W   =  2.25 \, GeV , \;
              T_3^e  = -0.5,        \;  
              Q_e   = -1, \; 
              S_W^2  = sin^2{\theta_{W}} = 0.232, \;
              C_W = cos{\theta_{W}}, \;
              C_V   =  2 T_3^e - 4 Q_e S_W^2, \;
              C_A   = -2 T_3^e, \;
              C_{L} = C_V - C_A , \;
              S_{avg} = 1/4 \,$ {\rm (spin averaging over the
                                 initial spin configuration)}, 
              $F_{ovl} = 128 \pi \alpha G_F^2 M_W^4 $.

\vskip 25pt
\begin{flushleft}
{\Large \bf A : The process $e^+ \, e^- \longrightarrow \nu
\bar{\nu} \, + \, \gamma$.}
\end{flushleft}
\vskip 10pt
\noindent
We label the particles by the following indices:
 $e^+ \Rightarrow 1 , \;
 e^- \Rightarrow 2 , \;
 \nu \Rightarrow 3 , \;
 \bar{\nu} \Rightarrow 4 , \;
 \gamma \Rightarrow 5$ .
\noindent
We have used the following abbreviations :
              $ P_{ij} = p_i.p_j, \;
          B_W = \frac{1}{(2P_{34} - M_Z^2)^2 + (M_Z{\Gamma}_Z)^2}, \;
            B =   2P_{34} - M_Z^2, \;
          W_3 = -(2P_{14} + M_W^2) , \;
          W_4 = -(2P_{23} + M_W^2) , \;
         W_{B3} = W_3^2 + (M_W{\Gamma}_W)^2, \;
         W_{B4} = W_4^2 + (M_W{\Gamma}_W)^2, \;
\epsilon{(ijkl)}= \epsilon_{\alpha \beta \gamma \delta}
P_i^{\alpha} P_j^{\beta} P_k^{\gamma} P_l^{\delta} $, where
$p_i$ is the momemtum of the $i-th$ particle.

\vskip 3pt
\noindent
In the following ${\bf T_{ij}} $ = $A_i A_j^ \dagger +$ H.C.,
where $A_i$ is the amplitude of the $i-th$ Feynman Diagram apart
from an overall  factor $F_{ovl}$ defined at the begining of
this appendix.  In this sub-section  we consider only the
diagrams (Fig.[11]) contributing to the cross section of the
process $e^+ \, e^- \longrightarrow \, \nu \, \bar{\nu} \, + \,
\gamma$.

\vskip 10pt
The relevant matrix element squared can be computed from the
following formulae:
\begin{flushleft}
$ {\bf T_{11}}= {\displaystyle \frac{B_W}{4C_W^4 P_{25}}} \Big
[(C_V^2+C_A^2) U_{11}+2 C_V C_A V_{11} \Big ] $,

\vskip 5pt
$ {\bf U_{11}}= P_{35}(P_{12} - 2P_{13} - P_{15}) +P_{13}
                (P_{15}+P_{25}) $ ,

$ {\bf V_{11}}=  P_{35}(P_{12}-P_{15})-P_{13} (P_{15}+P_{25}) $.
\end{flushleft}

\vskip 8pt
\begin{flushleft}
$        {\bf T_{12}} =  {\displaystyle{\frac{-B_W}{
4C_W^4P_{15}P_{25}}}} \Big [ 
(C_V^2+C_A^2) U_{12}+2 C_V C_A V_{12} \Big ]  $,
\vskip 5pt
$ {\bf U_{12}} = -2 P_{12}^2 (P_{13} + P_{23} - P_{35}) +
              2  P_{12} P_{13} (P_{15} + 2 P_{23} + P_{25} -
              P_{35}) - P_{12}(P_{15}+P_{25})(P_{35}-2P_{23}) $ \\
\hskip 34pt $+
(P_{13}P_{25}-P_{15}P_{23})(2P_{13}+P_{15}-2P_{23}-P_{25}) -
2P_{12}P_{23}P_{35}  $,

$ {\bf V_{12}} = 2 P_{12}^2 (P_{13} - P_{23}) -
2P_{12}P_{13}(P_{15}+2P_{25}) + P_{12}P_{15}(4P_{23}-P_{35}) +
P_{12}P_{25}(2P_{23}+P_{35}) $\\
\hskip 34pt $+ (P_{15}+P_{25})(P_{13}P_{25}-P_{15}P_{23}) $.
\end{flushleft}

\vskip 8pt
\begin{flushleft}
$        {\bf T_{13}} = -{\displaystyle{\frac{2B_W
C_L}{C_W^2P_{25}W_{B3}}}} (B W_3 
+ M_W M_Z {\Gamma}_W {\Gamma}_Z) 
 \Big [ P_{13}(P_{15}+P_{25}-P_{35}) \Big ] $.
\end{flushleft}

\vskip 7pt
\begin{flushleft}

$        {\bf T_{14}} = {\displaystyle{\frac{-B_W
C_{L}}{C_W^2P_{15}P_{25} W_{B4}}}} \Big [(B W_4 + M_W M_Z {\Gamma}_W 
{\Gamma}_Z) R_{14} 
               + (M_Z {\Gamma}_Z W_4 - M_W {\Gamma}_W B)
I_{14} \Big ] $,
\vskip 5pt
$        {\bf R_{14}}= P_{13}(P_{14}P_{25}-P_{12}P_{45}) -
P_{13}P_{24}(P_{15}+P_{25}) - P_{12}P_{24}(P_{35}-2P_{13}) +
P_{15}P_{23}P_{24} $,
 
$        {\bf I_{14}}= - \epsilon{(3125)}P_{24} +
         \epsilon{(4125)}P_{13} $.
\vskip 5pt
\end{flushleft}

\vskip 8pt
\begin{flushleft}

$   {\bf T_{15}} = {\displaystyle{\frac{B_WC_L}{C_W^2 P_{25} W_{B3}
W_{B4}}}} \Big[ \Big \{B (W_3 W_4 - M_W^2 {\Gamma}_W^2) + M_Z M_W
{\Gamma}_Z {\Gamma}_W (W_3 + W_4) \Big \} R_{15} $\\ 
\hskip 150pt $- \Big \{ B (W_3 + W_4) M_W
{\Gamma}_W -  M_Z {\Gamma}_Z (W_3 W_4 - M_W^2 {\Gamma}_W^2) \Big \}
I_{15} \Big] $,
\vskip 5pt
$   {\bf R_{15}} = (3P_{24}-P_{45})(P_{15}P_{23}
-P_{12}P_{35})-P_{13}P_{24}(P_{15}-2P_{23}+2P_{24}+2P_{25} -
2P_{45}-2P_{12}) $\\
\hskip 34pt $+ P_{14}P_{25}(P_{13}-P_{23}) - 
P_{13}P_{45}(P_{12}-3P_{25}+2P_{23}) + P_{12}P_{25}P_{34} $,

$   {\bf I_{15}} = -{\epsilon}{(3425)} \Big [ P_{12}+P_{15} \Big ] +
\epsilon{(4125)} \Big [ P_{13}-P_{23}-P_{35} \Big ] 
  - 2  \epsilon{(3125)} (P_{24}-P_{45}) + \epsilon{(3415)} P_{25} $
\end{flushleft}

\vskip 7pt
\begin{flushleft}
$        {\bf T_{22}} = {\displaystyle{\frac{B_W}{4C_W^4P_{15}}}} \Big
[(C_V^2+C_A^2) U_{22}+2 C_V C_A V_{22} \Big ] $,
\vskip 5pt
$        {\bf U_{22}}= (P_{12}-2P_{23}-P_{25})P_{35} +
(P_{15}+P_{25})P_{23} $,

$        {\bf V_{22}}= (P_{25}-P_{12})P_{35}+(P_{15}+P_{25})P_{23} $.
\end{flushleft}

\vskip 8pt
\begin{flushleft}

$        {\bf T_{23}} = {\displaystyle{\frac{-
B_WC_L}{C_W^2P_{15}P_{25}W_{B3}}}} \Big [(B W_3 + M_W M_Z {\Gamma}_W
{\Gamma}_Z) R_{23}  
               + (M_Z {\Gamma}_Z W_3 - M_W {\Gamma}_W B)
I_{23} \Big ] $,
\vskip 5pt
$        {\bf R_{23}}= P_{13}P_{25}(P_{14}-P_{24}) -
P_{15}P_{24}(P_{13}-P_{23}) + P_{12}P_{24}(2P_{13}-P_{35}) -
P_{12}P_{13}P_{45} $,

$        {\bf I_{23}}=  \epsilon{(3125)}P_{24} - \epsilon{(4125)}
P_{13} $.
\end{flushleft}

\vskip 8pt
\begin{flushleft}
$        {\bf T_{24}} =
{\displaystyle{\frac{-2B_WC_L}{C_W^2P_{15}W_{B4}}}}(B W_4 + M_W M_Z
{\Gamma}_W {\Gamma}_Z) \Big [P_{35}(P_{12}-P_{23}-P_{25}) \Big ]$.

\end{flushleft}

\vskip 9pt
\begin{flushleft}
$        {\bf T_{25}} = {\displaystyle{\frac{-B_W
C_{L}}{C_W^2P_{15}W_{B3}W_{B4}}}} \Big [ \Big \{ B (W_3 W_4 - M_W^2
{\Gamma}_W^2) + M_Z M_W {\Gamma}_Z {\Gamma}_W (W_3 + W_4) \Big \}
R_{25} $\\
\hskip 150pt $- \Big \{ B (W_3 + W_4) M_W {\Gamma}_W - 
M_Z {\Gamma}_Z (W_3  
                 W_4 - M_W^2 {\Gamma}_W^2)  \Big \} I_{25}\Big ] $,
\vskip 5pt
$       {\bf R_{25}}= -P_{12}P_{13}(P_{34}-P_{45}) -
P_{12}P_{35}(P_{24}-P_{34}-P_{45}) +
P_{13}P_{14}(P_{23}-2P_{24}-P_{25}) $\\
\hskip 34pt $+ P_{13}P_{24}(P_{13}+P_{15}+P_{25}-P_{35}+P_{45}) -
P_{14}P_{35}(P_{23}-P_{24}+P_{25}) -
P_{15}P_{24}(P_{23}+P_{34}+3P_{35}) $,

$       {\bf I_{25}}=   \epsilon{(3415)}(P_{13}+P_{24}) +
\epsilon{(4125)}(P_{13}+P_{35}) - \epsilon{(3125)}P_{24} -
\epsilon{(3412)}P_{35} $.
\end{flushleft}

\vskip 9pt
\begin{flushleft}
$        {\bf T_{33}} = {\displaystyle{\frac{4}{P_{25}W_{B3}}}}
\Big [ P_{13}(P_{15}+P_{25}-P_{35}) \Big ] $.
\end{flushleft}

\vskip 8pt
\begin{flushleft}
$        {\bf T_{34}} =
{\displaystyle{\frac{4}{P_{15}P_{25}W_{B3}W_{B4}}}} \Big [(W_3
W_4 + M_W^2 {\Gamma}_W^2) R_{34} - M_W {\Gamma}_W (W_3 - W_4)
I_{34} \Big ] $,
\vskip 5pt
$        {\bf R_{34}}=
P_{12}P_{13}(2P_{12}-2P_{15}-2P_{23}-3P_{25}+P_{35}) -
P_{12}P_{35}(P_{12}-P_{23}-P_{25})$\\
\hskip 34pt $- (P_{13}-P_{23}-P_{25})(P_{13}P_{25}-P_{15}P_{23}) +
P_{12}P_{15}P_{23} $,

$        {\bf I_{34}}= -
\epsilon{(3125)}(P_{12}+P_{13}-P_{23}-P_{25})$.
\end{flushleft}

\vskip 9pt
\begin{flushleft}
$        {\bf T_{35}} =
{\displaystyle{\frac{-4}{P_{25}W_{B3}W_{B4}}}}
\Big [W_4 R_{35} - M_W {\Gamma}_W I_{35}\Big ] $,
\vskip 5pt
$        {\bf R_{35}}=
-P_{12}P_{35}(3P_{12}+P_{13}-P_{15}-3P_{23}-3P_{25}+P_{35}) +
P_{12}P_{13}(4P_{23}+P_{25}) +
P_{15}P_{23}(3P_{12}-3P_{13}-P_{15}-3P_{23} $ \\
\hskip 34pt $-3P_{25}+P_{35}) +
P_{13}P_{25}(P_{13}+3P_{15}-5P_{23}+3P_{25}-3P_{35}) -
4P_{13}P_{23}(P_{23}-P_{35}) $,

$        {\bf I_{35}}=
-(3P_{12}+P_{13}-P_{15}-3P_{23}-3P_{25}+P_{35}).
\epsilon{(3125)} $
\end{flushleft}

\vskip 7pt
\begin{flushleft}
$        {\bf T_{44}} = {\displaystyle{\frac{-4}{P_{15}W_{B4}}}}
\Big [P_{35}(P_{12}-P_{23}-P_{25}) \Big ] $.
\end{flushleft}

\vskip 8pt
\begin{flushleft}
$        {\bf T_{45}} =
{\displaystyle{\frac{8}{P_{15}W_{B3}W_{B4}}}}
\Big [W_3 R_{45} - M_W {\Gamma}_W I_{45} \Big ] $,
\vskip 5pt
$        {\bf R_{45}}=
-P_{12}P_{13}(P_{12}-P_{13}-2P_{15}-P_{23}-2P_{25}+P_{35}) -
P_{12}P_{15}(P_{23}+P_{35}) -
P_{13}P_{23}(2P_{13}+2P_{15}+P_{25}-2P_{35})$\\
\hskip 34pt $- P_{13}P_{25}(P_{13}+P_{15}+P_{25}-2P_{35}) +
P_{15}P_{23}(P_{23}+P_{25}+2P_{35}) -
P_{35}(P_{12}P_{35}-2P_{15}P_{25}) $,

$        {\bf I_{45}}= -(P_{12}+P_{13}-P_{23}-P_{25})
\epsilon{(3125)} $.
\end{flushleft}

\vskip 8pt
\begin{flushleft}
$        {\bf T_{55}} = {\displaystyle{\frac{-4}{W_{B3}W_{B4}}}}
\Big [ U_{55} \Big ] $,
\vskip 5pt
$        {\bf U_{55}}=
-P_{12}P_{23}(P_{12}+6P_{13}+5P_{15}-P_{23}-P_{25}+4P_{35}) +
P_{12}P_{35}(3P_{12}+P_{13}-3P_{15}-3P_{25}-P_{35}) $ \\
\hskip 34pt $-P_{13}P_{25}(P_{12}+3P_{13}+3P_{15}-8P_{23}
+3P_{25}-7P_{35}) - P_{13}P_{23}(P_{13}-5P_{15}-7P_{23}+3P_{35})$ \\
\hskip 34pt $+ 2P_{15}P_{23}(P_{15}+3P_{23}+3P_{25}+P_{35}) +
4P_{15}P_{25}P_{35} $.
\end{flushleft}
\begin{eqnarray*}
        T(\nu \bar{\nu}) &=& 3(T_{11} + T_{12} + T_{22}) +
T_{13} + T_{14} + T_{15} \\ &+& T_{23} + T_{24} + T_{25} +
T_{33} + T_{34} + T_{35} \\ &+& T_{44} + T_{45} + T_{55}
\end{eqnarray*}
\vskip 10pt
The differential cross section is given by
\[ d \sigma = F_{ovl} S_{avg} \frac{T(\nu
\bar{\nu})}{64 E_{CM}^2 \pi ^5}  \delta ^4 \Big ( p_1+p_2- \sum_{i=3}^5
p_i \Big ) \prod_{i=3}^5 \frac{d^3p_i}{2E_i} \] Other cross
sections are obtained by replacing the $T$-factor in the above
formula by the appropriate expressions calculated in the
following appendices.
\vskip 20pt
\begin{flushleft}
{\Large \bf B : The process $e^+ \, e^- \longrightarrow
\tilde{\nu}
\tilde{\bar{\nu}} \, + \, \gamma$.}
\end{flushleft}
\vskip 10pt
\noindent
In this subsection we consider the diagrams (Fig.[12])
contributing to the cross section of the process $e^+e^-
\longrightarrow \tilde{\nu} \tilde{\bar{\nu}} + \gamma $. We
define for this subsection:
\[  B_W = \frac{1}{ \Big \{2(\MSNUSQ + P_{34}) - M_Z^2 \Big \}^2 +
(M_ZG_Z)^2}. \]
\noindent
We label the particles by the following indices: $\tilde{\nu}
\Rightarrow 3 , \tilde{\bar{\nu}} \Rightarrow 4$ , while the
indices 1,2 and 5 have the same meanining as in the previous
subsection.
\vskip 8pt
\begin{flushleft}
The convention for the $T_{ij}$-s in this subsection is the same
as in the last subsection for the process under consideration.
\end{flushleft}

\vskip 7pt
\begin{flushleft}
$ {\bf T_{11}} =  {\displaystyle{\frac{3 B_W }{8C_W^4P_{25}}}}
(\CVSQ+\CASQ)
\Big [ (P_{13}-P_{14})(P_{35}-P_{45}) + P_{15}(P_{34}-\MSNUSQ) \Big ] $.
\end{flushleft}

\vskip 10pt
\begin{flushleft}
$ {\bf T_{12}} = {\displaystyle{\frac{-3
B_W}{8C_W^4P_{15}P_{25}}}} (\CVSQ+\CASQ) U_{12} $,

$ {\bf U_{12}} = P_{12} \Big  \{
2(P_{12}-P_{15}-P_{25})(m_{\tilde{\nu}}^2-P_{34}) -
(P_{13}-P_{14}) (2P_{23}-2P_{24}-P_{35}+P_{45}) +
(P_{23}-P_{24})(P_{35}-P_{45}) \Big \} $\\
\hskip 34pt $ + P_{15} 
\Big \{ P_{23}(P_{13}-P_{14}-P_{23}+2P_{24}) 
- P_{24}(P_{13}-P_{14}+P_{24})
\} - P_{25}  \{ P_{13}(P_{13}-2P_{14}-P_{23}+P_{24}) $\\ 
\hskip 34pt $ + P_{14}(P_{14}+P_{23}-P_{24}) \Big \} $.
\end{flushleft}

\vskip 15pt
\begin{flushleft}
$ {\bf T_{13}} = -\displaystyle{\frac{B_W C_L}{2C_W^2P_{25}}}
(\MZSQ-2\MSNUSQ-2P_{34})\sum_{a=1}^2
\frac{\VASQ}{(\MSNUSQ-\MASQ-2P_{14})} \Big [U_{13} \Big ] $ , \\

$ {\bf U_{13}} = -P_{12}(2P_{25}-3P_{35}) +
P_{13}(3P_{25}-4P_{35}) + P_{15}(2\MSNUSQ-3P_{23}) $ \\ where
${\cal V}_{a1}$, $a$=1,2 are the mixing factors corresponding to
the two charginos.
\end{flushleft}

\vskip 15pt
\begin{flushleft}
$ {\bf T_{14}} =  -\displaystyle{\frac{B_W
C_L}{2C_W^2P_{15}P_{25}}} (\MZSQ-2\MSNUSQ-2P_{34})\sum_{a=1}^2
\frac{\VASQ}{(\MSNUSQ-\MASQ-2P_{23})} \Big [U_{14} \Big ] $ , \\

$  {\bf U_{14}} = 2( \MSNUSQ
P_{12}-P_{13}P_{23})(P_{12}-P_{15}-P_{25}) +
P_{12}P_{35}(2P_{13}+2P_{23}-P_{25}) -
P_{13}P_{25}(2P_{13}-P_{25}) $\\
\hskip 34pt $- P_{15}P_{23}(2P_{23}+P_{25}) - 2P_{12}P_{13}P_{23} $.
\end{flushleft}

\vskip 15pt
{\begin{flushleft} $ {\bf T_{15}} =  -\displaystyle{\frac{B_W
C_L}{C_W^2P_{25}}} (\MZSQ-2\MSNUSQ-2P_{34})\sum_{a=1}^2
\frac{\VASQ}{(\MSNUSQ-\MASQ-2P_{14})(\MSNUSQ-\MASQ-2P_{23})} \Big
[2U_{15}-\MASQ W_{15} \Big ] $ , \\

$ {\bf U_{15}} = P_{23} \Big \{ \MSNUSQ(P_{15}-P_{12}) -
P_{25}(P_{12}-3P_{13}-P_{15}) + 2P_{13}(P_{23}-P_{35}) \Big \}
$,

$ {\bf W_{15}} = P_{12}(P_{25}-P_{35}) - P_{13}P_{25} +
P_{15}P_{23} $.
\vskip 5pt
\end{flushleft}

\vskip 15pt
\begin{flushleft}
$ {\bf T_{22}} = \displaystyle{\frac{3 B_W}{4C_W^4P_{15}}}
(\CVSQ+\CASQ)
\Big [ (P_{23}-P_{24})(P_{35}-P_{45}) + P_{25}(P_{34}-\MSNUSQ) \Big ] $.
\end{flushleft}

\vskip 15pt
\begin{flushleft}
$ {\bf T_{23}} =  - \displaystyle{\frac{-B_W
C_L}{2C_W^2P_{15}P_{25}}} (\MZSQ-2\MSNUSQ-2P_{34})\sum_{a=1}^2
\frac{\VASQ}{(\MSNUSQ-\MASQ-2P_{14})} \Big [U_{23} \Big ] $ , \\

$ {\bf U_{23}} = -P_{12} \Big \{ 2 \MSNUSQ
(P_{12}-P_{15}-P_{25}) - 2P_{13}(2P_{23}-P_{35}) +
P_{15}(2P_{25}-P_{35}) + 2P_{35}(P_{23}-P_{25})
\Big \} $\\
\hskip 34pt 
$+ (P_{13}P_{25}-P_{15}P_{23})(2P_{13}+P_{15}-2P_{23}-2P_{25}) $.
\vskip 5pt
\end{flushleft}

\vskip 15pt
\begin{flushleft}
$ {\bf T_{24}} =  \displaystyle{\frac{B_W C_L}{2C_W^2P_{15}}}
(\MZSQ-2\MSNUSQ-2P_{34})\sum_{a=1}^2
\frac{\VASQ}{(\MSNUSQ-\MASQ-2P_{23})} \Big [U_{24} \Big ] $ , \\

$ {\bf U_{24}} = P_{25}(P_{13}-2 \MSNUSQ) - P_{35}(P_{12}-4P_{23}) -
P_{15}P_{23} $.
\vskip 5pt
\end{flushleft}

\vskip 15pt
\begin{flushleft}
$ {\bf T_{25}} =  \displaystyle{\frac{B_W C_L}{C_W^2P_{15}}}
(\MZSQ-2\MSNUSQ-2P_{34})\sum_{a=1}^2 
\frac{\VASQ}{(\MSNUSQ-\MASQ-2P_{14})(\MSNUSQ-\MASQ-2P_{23})} \Big
[2U_{25}-\MASQ W_{25} \Big ] $ , \\

$ {\bf U_{25}} = \MSNUSQ \Big \{ P_{12}(P_{12}-P_{13}-P_{15}-P_{25}) +
P_{25}(P_{13}+P_{15}) \Big \} -
P_{23}(2P_{13}+P_{15}-2P_{35}) $\\
\hskip 34pt $(P_{12}-P_{13}-P_{15}) $,

$ {\bf W_{25}} = P_{12}(P_{15}-P_{35}) + P_{13}P_{25} - P_{15}P_{23} $.
\vskip 5pt
\end{flushleft}

\vskip 15pt
\begin{flushleft}
$ {\bf T_{33}} =  \displaystyle{\frac{4}{P_{25}}}
\sum_{a=1}^2 \sum_{b=1}^2 
\frac{\VASQ \VBSQ}{(\MSNUSQ-\MASQ-2P_{14})(\MSNUSQ-\MBSQ-2P_{14})} \Big
[2P_{14}P_{45}- \MSNUSQ P_{15} \Big ] $ . \\
\end{flushleft}

\begin{flushleft}
$ {\bf T_{34}} =  \displaystyle{\frac{4}{P_{15}P_{25}}}
\sum_{a=1}^2 \sum_{b=1}^2 
\frac{\VASQ \VBSQ}{(\MSNUSQ-\MASQ-2P_{14})(\MSNUSQ-\MBSQ-2P_{23})} \Big
[ U_{34} \Big ] $,

$ {\bf U_{34}} = P_{12} \Big \{- \MSNUSQ (P_{12}-P_{15}-P_{25}) +
P_{13}(2P_{23}-P_{35}) - P_{35}(P_{23}-P_{25})  \Big \}$\\
\hskip 34pt $ +(P_{13}P_{25}-P_{15}P_{23})(P_{13}-P_{23}-P_{25}) $.
\end{flushleft}

\vskip 15pt
\begin{flushleft}
$ {\bf T_{35}} =  \displaystyle{\frac{4}{P_{25}}}
\sum_{a=1}^2 \sum_{b=1}^2 
\frac{\VASQ
\VBSQ}{(\MSNUSQ-\MASQ-2P_{14})(\MSNUSQ-\MBSQ-2P_{14})
(\MSNUSQ-\MBSQ-2P_{23})}
\Big [ 2U_{35}- \MBSQ W_{35} \Big ] $,

$ {\bf U_{35}} = P_{23} \Big \{\MSNUSQ (P_{12}-P_{15}) -
2P_{14}(P_{24}-P_{45}) \Big \} $,

$ {\bf W_{35}} = -P_{12}P_{45}-P_{14}P_{25}+P_{15}P_{24} $.
\end{flushleft}

\vskip 15pt
\begin{flushleft}
$ {\bf T_{44}} =  \displaystyle{\frac{4}{P_{15}}}
\sum_{a=1}^2 \sum_{b=1}^2 
\frac{\VASQ
\VBSQ}{(\MSNUSQ-\MASQ-2P_{23})(\MSNUSQ-\MBSQ-2P_{23})}
\Big [ 2P_{23}P_{35} - \MSNUSQ P_{25} \Big ] $ . \\
\end{flushleft}

\vskip 15pt
\begin{flushleft}
$ {\bf T_{45}} =  \displaystyle{\frac{4}{P_{15}}}
\sum_{a=1}^2 \sum_{b=1}^2 
\frac{\VASQ \VBSQ}{(\MSNUSQ-\MASQ-2P_{23})(\MSNUSQ-\MBSQ-2P_{23})
(\MSNUSQ-\MBSQ-2P_{14})} \Big 
[ 2U_{45}- \MBSQ W_{45} \Big ] $ , \\

$ {\bf U_{45}} = \MSNUSQ \Big \{ P_{12}(P_{12}-P_{13}-P_{15}-P_{25}) +
P_{25}(P_{13}+P_{15}) \Big \} - 2P_{13}P_{23}
(P_{12}-P_{13}-P_{15}+P_{35}) $\\
\hskip 34pt $+ 2P_{23}P_{35}(P_{12}-P_{15}) $,

$ {\bf W_{45}} = -P_{12} P_{35}+P_{13} P_{25}-P_{15} P_{23} $.
\vskip 5pt
\end{flushleft}

\vskip 15pt
\begin{flushleft}
$ {\bf T_{55}} = \displaystyle{\frac{8}{P_{15}}}
\sum_{a=1}^2 \sum_{b=1}^2 
\frac{\VASQ \VBSQ \Big [ U_{55}-2(\MASQ+\MBSQ) W_{55} 
+ \MASQ \MBSQ P_{12}
\Big ]}{(\MSNUSQ-\MASQ-2P_{14})(\MSNUSQ-\MASQ-2P_{23})
(\MSNUSQ-\MBSQ-2P_{14})(\MSNUSQ-\MBSQ-2P_{23})} $, \\

$ {\bf U_{55}} = \MSNUSQ \Big \{ P_{12}(\MSNUSQ -4P_{23}+2P_{35}) -
2P_{13}P_{25} + 2P_{15}(P_{23}-P_{25}) \Big \} $\\
\hskip 34pt$ +4P_{23}(P_{23}-P_{35})(P_{12}-P_{15}) 
+ 4P_{13}P_{23}P_{25} $, \\

$ {\bf W_{55}} = P_{23} (-P_{12}+P_{13}+P_{15}) $.
\end{flushleft}

\[ T(\tilde{\nu} \tilde{\bar{\nu}})=\sum_{{\begin{array}{rr}
           i,j = 1 \\
             j  \geq  1
         \end{array}}}^5 T_{ij} \]
\vskip 10pt

\vskip 20pt
\begin{flushleft}
{\Large \bf C : The process $e^+ \, e^- \longrightarrow
\widetilde{N}_i \widetilde{N}_j \, + \, \gamma$.}
\end{flushleft}
\vskip 10pt
\noindent
In this subsection we give the formula for the cross section of
the process $e^+e^- \longrightarrow \widetilde{N}_i
\widetilde{N}_j + \gamma $. The contributing Feynman diagrams
are as in Fig.[13]. We define for this subsection:
\begin{eqnarray*}
 S &=& \MI2 + \MJ2 + 2P_{34}- \MZSQ \\
 B_W &=& \frac{1}{(\MI2 + \MJ2 + 2P_{34} - \MZSQ)^2 +
M_Z^2G_Z^2 }
\end{eqnarray*}
\noindent
In this subsection we label the particles by the following
indices: 3 and 4 stand for $\NI$ and $\NJ$ respectively where
$i,j$=1,2 and $i \leq j$. The labels 1,2,5 have the same meaning
as in the previous subsection.
\begin{flushleft}
The convention for the $T_{ij}$-s in this subsection is the same
as in the last subsection for the process under consideration.
\end{flushleft}


\vskip 25pt
\begin{flushleft}
$ {\bf T_{11}} = -{\displaystyle{\frac{(\CASQ+\CVSQ) \GASQ}{2
P_{25}}}} \Big [m_i m_j P_{15}-P_{14}P_{35}-P_{13}P_{45} \Big ]
$

\vskip 15pt
$ {\bf T_{12}} = -{\displaystyle{\frac{(\CASQ+\CVSQ) \GASQ}{2
P_{15}P_{25}}}} \Big [ (P_{12}-P_{15}-P_{25}) (2m_im_j
P_{12}-P_{13}P_{24}-P_{14}P_{23}) - P_{12} \Big \{
P_{13}P_{24}+P_{14}P_{23} $ \\ 
f\hskip 110pt $-P_{35}(P_{14}+P_{24})-P_{45}(P_{13}+P_{23}) \Big
\} - 2(P_{15}P_{23}P_{24}+P_{13}P_{14}P_{25}) \Big ] $

\vskip 15pt
$ {\bf T_{13}} = -{\displaystyle{\frac{2C_L
G_A}{P_{25}(\MJ2-2P_{14}-\MSEL2)}}} \Big [ m_i m_jP_{15}-2P_{14}P_{35}
\Big ] $


\vskip 10pt
$ {\bf T_{14}} = {\bf -T_{13}}\left \{ 
   \begin{array} {ll}
      i  \leftrightarrow  j \\
      3  \leftrightarrow  4
   \end{array}
 \right \} $

\vskip 15pt
$ {\bf T_{15}} = {\displaystyle{\frac{2C_L G_A}{
P_{15}P_{25}(\MI2-2P_{23}-\MSEL2)}}} \Big [
(P_{12}-P_{15}-P_{25})(m_im_jP_{12}-P_{14}P_{23}) -
P_{23}(P_{12}P_{14}+P_{15}P_{24}-P_{12}P_{45}) $ \\
\hskip 150pt $-P_{14}(P_{13}P_{25}-P_{12}P_{35}) \Big ] $

\vskip 10pt
$ {\bf T_{16}} = {\bf -T_{15}}\left \{ 
   \begin{array} {ll}
      i  \leftrightarrow  j \\
      3  \leftrightarrow  4
   \end{array}
 \right \} $

\vskip 15pt
$ {\bf T_{17}} = {\displaystyle{\frac{C_L
G_A}{P_{25}(\MI2-2P_{23}-\MSEL2)(\MJ2-2P_{14}-\MSEL2)}}} \Big [
m_im_j  \Big \{ P_{12}(-2P_{12}+2P_{15}-2P_{23}+2P_{24}+P_{35}-P_{45}) 
$\\ \hskip 150pt $+ P_{15}(P_{23}-P_{24}) -
P_{25}(P_{13}-P_{14}) \Big \} + 2P_{14} \Big
\{P_{23}(2P_{12}-P_{15}+2P_{23} $\\
\hskip 150pt $-2P_{24}-2P_{35}+P_{45}) +
P_{25}(\MI2+P_{13}-P_{34}) - P_{35}(P_{12}-P_{24}) \Big \} \Big ] $

\vskip 10pt
$ {\bf T_{18}} = {\bf -T_{17}}\left \{ 
   \begin{array} {ll}
      i  \leftrightarrow  j \\
      3  \leftrightarrow  4
   \end{array}
 \right \} $

$ {\bf T_{22}} = {\bf T_{11}}(1\leftrightarrow 2) $

\vskip 15pt
$ {\bf T_{23}} = {\displaystyle{\frac{2C_L
G_A}{P_{15}P_{25}(\MI2-2P_{14}-\MSEL2)}}} \Big [
(P_{12}-P_{15}-P_{25})(m_im_jP_{12}-P_{14}P_{23}) -
P_{23}(P_{12}P_{14}-P_{12}P_{45}+P_{15}P_{24}) $\\
\hskip 150pt $ - P_{14}(P_{13}P_{25}-P_{12}P_{35}) \Big ] $

\vskip 10pt
$ {\bf T_{24}} = {\bf -T_{23}}\left \{ 
   \begin{array} {ll}
      i  \leftrightarrow  j \\
      3  \leftrightarrow  4
   \end{array}
 \right \} $

\vskip 15pt
$ {\bf T_{25}} = {\displaystyle{\frac{2C_L
G_A}{P_{15}(\MI2-2P_{23}-\MSEL2)}}} \Big [ m_im_jP_{25} 
- 2P_{23}P_{45} \Big ] $

\vskip 10pt
$ {\bf T_{26}} = {\bf -T_{25}}\left \{ 
   \begin{array} {ll}
      i  \leftrightarrow  j \\
      3  \leftrightarrow  4
   \end{array}
 \right \} $

\vskip 15pt
$ {\bf T_{27}} = {\displaystyle{\frac{C_L
G_A}{P_{15}(\MI2-2P_{23}-\MSEL2)(\MJ2-2P_{14}-\MSEL2) }}} \Big [
m_im_j \Big \{
P_{12}(-2P_{12}+2P_{13}-2P_{14}+2P_{25}-P_{35}+P_{45}) $ \\
\hskip 150pt $+ P_{15}(P_{23}-2P_{24}) - P_{25}(P_{13}-P_{14}) \Big \} +
2P_{23} \Big \{ P_{14}(2P_{12}-2P_{13}+2P_{14} $ \\
\hskip 150pt $-P_{25}+P_{35}-2P_{45}) +
P_{15}(\MJ2+P_{24}-P_{34}) - P_{45}(P_{12}-P_{13}) \Big \} \Big ] $

\vskip 10pt
$ {\bf T_{28}} = {\bf -T_{27}}\left \{ 
   \begin{array} {ll}
      i  \leftrightarrow  j \\
      3  \leftrightarrow  4
   \end{array}
 \right \} $

\vskip 15pt
$ {\bf T_{33}} = {\displaystyle{\frac{16P_{14}P_{35}}
{P_{25}(\MJ2-2P_{14}-\MSEL2)^2}}} $

\vskip 15pt
$ {\bf T_{34}} = {\displaystyle{\frac{16m_im_jP_{15}}
{P_{25}(\MI2-2P_{13}-\MSEL2)(\MJ2-2P_{14}-\MSEL2)}}} $

\vskip 15pt
$ {\bf T_{35}} = {\displaystyle{\frac{16}
{P_{15}P_{25}(\MI2-2P_{23}-\MSEL2)(\MJ2-2P_{14}-\MSEL2)}}} \Big [
P_{14}P_{23}(2P_{12}-P_{15}-P_{25}) 
+ P_{14}(P_{13}P_{25}-P_{12}P_{35}) $ \\
\hskip 250pt $+ P_{23}(P_{15}P_{24}-P_{12}P_{45}) \Big ] $

\vskip 15pt
$ {\bf T_{36}} = {\displaystyle{\frac{16 m_im_j 
P_{12}(P_{12}-P_{15}-P_{25})}
{P_{15}P_{25}(\MJ2-2P_{24}-\MSEL2)(\MJ2-2P_{14}-\MSEL2)}}} $

\vskip 15pt
$ {\bf T_{37}} = {\displaystyle{\frac{-16P_{14}}
{P_{25}(\MI2-2P_{23}-\MSEL2)(\MJ2-2P_{14}-\MSEL2)^2}}} \Big [
P_{23}(2P_{12}-P_{15}+2P_{23}-2P_{24}-2P_{35}+P_{45}) $ \\
\hskip 250pt $+ P_{25}(\MI2+P_{13}-P_{34}) 
- P_{35}(P_{12}-P_{24}) \Big ] $

\vskip 15pt
$ {\bf T_{38}} = {\displaystyle{\frac{-8m_im_j}
{P_{25}(\MI2-2P_{13}-\MSEL2)(\MJ2-2P_{14}-\MSEL2)(\MJ2-2P_{24}-\MSEL2)}}}
\Big [ P_{12}(2P_{12}-2P_{15}-2P_{23}+2P_{24} $ \\
\hskip 250pt $ +P_{35}-P_{45}) +
P_{15}(P_{23}-P_{24}) - P_{25}(P_{13}-P_{14}) \Big ] $

\vskip 10pt
$ {\bf T_{44}} = {\bf T_{33}}\left \{ 
   \begin{array} {ll}
      i  \leftrightarrow  j \\
      3  \leftrightarrow  4
   \end{array}
 \right \} $

\vskip 10pt
$ {\bf T_{45}} = {\bf T_{36}}\left \{ 
   \begin{array} {ll}
      i  \leftrightarrow  j \\
      3  \leftrightarrow  4
   \end{array}
 \right \} $

\vskip 10pt
$ {\bf T_{46}} = {\bf T_{35}}\left \{ 
   \begin{array} {ll}
      i  \leftrightarrow  j \\
      3  \leftrightarrow  4
   \end{array}
 \right \} $

\vskip 10pt
$ {\bf T_{47}} = {\bf T_{38}}\left \{ 
   \begin{array} {ll}
      i  \leftrightarrow  j \\
      3  \leftrightarrow  4
   \end{array}
 \right \} $

\vskip 10pt
$ {\bf T_{48}} = {\bf T_{37}}\left \{ 
   \begin{array} {ll}
      i  \leftrightarrow  j \\
      3  \leftrightarrow  4
   \end{array}
 \right \} $

\vskip 15pt
$ {\bf T_{55}} =
{\displaystyle{\frac{16P_{23}P_{45}}{P_{15}(\MI2-2P_{23}-\MSEL2)^2}}} $

\vskip 15pt
$ {\bf T_{56}} =
{\displaystyle{\frac{16m_im_jP_{25}}{P_{15}(\MI2-2P_{23}-\MSEL2) 
(\MJ2-2P_{24}-\MSEL2)}}} $

\vskip 15pt
$ {\bf T_{57}} 
= {\displaystyle{\frac{-16P_{23}}{P_{15}(\MI2-2P_{23}-\MSEL2)^2
(\MJ2-2P_{14}-\MSEL2)}}} \Big [
(P_{12}-P_{13}+P_{14})(2P_{14}-P_{45}) $\\
\hskip 250pt $- P_{14}(P_{25}-P_{35}+P_{45}) 
+ P_{15}(\MJ2+P_{24}-P_{34})
\Big ] $ 

\vskip 15pt
$ {\bf T_{58}} =
{\displaystyle{\frac{-8m_im_j}{P_{15}(\MI2-2P_{13}-\MSEL2)
(\MI2-2P_{23}-\MSEL2)(\MJ2-2P_{24}-\MSEL2)}}} \Big [
(P_{13}-P_{14})(2P_{12}-P_{25}) $ \\
\hskip 250pt $+ P_{12} \Big \{
2(P_{12}-P_{25})-(P_{35}-P_{45}) \Big \} + P_{15}(P_{23}-P_{24})
\Big ] $

\vskip 10pt
$ {\bf T_{66}} = {\bf T_{55}}\left \{ 
   \begin{array} {ll}
      i  \leftrightarrow  j \\
      3  \leftrightarrow  4
   \end{array}
 \right \} $

\vskip 10pt
$ {\bf T_{67}} = {\bf T_{58}}\left \{ 
   \begin{array} {ll}
      i  \leftrightarrow  j \\
      3  \leftrightarrow  4
   \end{array}
 \right \} $

\vskip 10pt
$ {\bf T_{68}} = {\bf T_{57}}\left \{ 
   \begin{array} {ll}
      i  \leftrightarrow  j \\
      3  \leftrightarrow  4
   \end{array}
 \right \} $

\vskip 15pt
$ {\bf T_{77}} =
{\displaystyle{\frac{-16P_{14}P_{23}}{(\MI2-2P_{23}-\MSEL2)^2
\: (\MJ2-2P_{14}-\MSEL2)^2}}} \Big [ \MI2 + \MJ2 -
2(P_{12}-P_{13}+P_{14}+P_{23}-P_{24}+P_{34}) \Big ] $ 

\vskip 15pt
$ {\bf T_{78}} = {\displaystyle{\frac{16m_im_jP_{12} \: \Big [
\MI2+\MJ2+2(P_{12}-P_{34}) \Big ]}{(\MI2-2P_{23}-\MSEL2) 
\: (\MI2-2P_{13}-\MSEL2) \: (\MJ2-2P_{14}-\MSEL2) \:
(\MJ2-2P_{24}-\MSEL2)}}} $

\vskip 10pt
$ {\bf T_{88}} = {\bf T_{77}}\left \{ 
   \begin{array} {ll}
      i  \leftrightarrow  j \\
      3  \leftrightarrow  4
   \end{array}
 \right \} $

\begin{eqnarray*}
 {\bf T_{st}} &=& T_{13} - T_{14} + T_{15} - T_{16} + T_{17} -
T_{18} \\ &+& T_{23} - T_{24} + T_{25} - T_{26} + T_{27} -
T_{28} \\ {\bf T_{t1}} &=&
T_{33}+T_{44}+T_{55}+T_{66}+T_{77}+T_{88}+T_{35}+T_{37}+T_{46}
+T_{48}+T_{57} +T_{68}\\ {\bf T_{t2}} &=& T_{34} + T_{36} +
T_{38} + T_{45} + T_{47} + T_{56} + T_{58} + T_{67} + T_{78}
\end{eqnarray*}
\vskip 5pt
\begin{eqnarray*}
  T(\widetilde{N}_i \widetilde{N}_j) &=& B_W(T_{11}+T_{12}+T_{22}) \\
             &+& \sum_{h=L,R} {\Big [B_W S \Big \{(a_i)_h(a_j)_h 
T_{st} \Big \} + \Big \{ (a_i)_h (a_j)_h \Big \} ^2 T_{t1} +
C_{ij} S_{ij}
\Big \{ (a_i)_h (a_j)_h \Big \} ^2 T_{t2} \Big ]}
\end{eqnarray*}
\end{flushleft}
where 
\begin{eqnarray*}
G_A &=& N_{i3}^ \prime N_{j3}^\prime - N_{i4}^ \prime N_{j4}^ \prime \\
(a_i)_L &=& (0.5 + S_W^2) N_{i2}^ \prime - S_WC_W N_{i1}^ \prime \\
(a_i)_R &=&  - S_W^2 N_{i2}^ \prime + S_WC_W N_{i1}^ \prime \\
(a_j)_L &=& (0.5 + S_W^2) N_{j2}^ \prime - S_WC_W N_{j1}^ \prime \\
(a_j)_R &=&  - S_W^2 N_{j2}^ \prime + S_WC_W N_{j1}^ \prime \\
\end{eqnarray*}
where the matrix $N^\prime$ diagonalises the 4$\times$4
neutralino mass matrix following the convention of Haber and
Kane[1].
\vskip 3pt
The Chiral Rotation Factor ($C_{ij}$) and the Fermi  statistics
Factor ($S_{ij}$) are defined as follows:
\begin{eqnarray*}
 C_{ij} &=& +1 \; \;
\;for \; 
\;m_{\tilde{N}_1},m_{\tilde{N}_2} > 0 \\ 
       &=& -1 \; \; \; for \; \; either \; \; m_{\tilde{N}_1} \; or \;
              m_{\tilde{N}_2} < 0 \\ 
 S_{ij} &=& +1 \; \;
\; for \; \; 
i \neq j \\ 
       &=& -1 \; \; \; for \; \; i=j
\end{eqnarray*}
\vskip 10pt

\newpage

\def \MSNU{m_{\tilde{\nu}}}
\def \GLUM{m_{\tilde{g}}}
\def \tanbeta{{\rm tan}\beta}
\def \MSELL{m_{{\tilde e}_L}}
\def \MSELR{m_{{\tilde e}_R}}

\begin{flushleft}
{\Large \bf Table Captions}\\
\end{flushleft}

\vskip 10pt
\begin{flushleft}
{\bf Table-I :}
\end{flushleft}

The comparison of the response of the signal to two sets of cuts
{\bf A}[6] and {\bf B}[15] at $\sqrt{s}$=190 GeV where {\rm Cut}
{\rm {\bf A}}  $\equiv$ 5$<E_{\gamma}<$60 GeV,
$40^\circ<\theta{\gamma}<140^\circ$; {\rm Cut}  {\rm {\bf B}} $
\equiv$  1$<E_{\gamma}<$47.5 GeV,
18$^\circ<\theta{\gamma}<$162$^\circ$, $p_{T_{\gamma}}>$6.175
GeV.  Other fixed values of the SUSY parameters used are
$(\mu,\GLUM,\tanbeta)$=($-$300 GeV, 200 GeV, 10) and
$\MSELL=\MSELR$.  The SM background with Cut {\bf A}({\bf B}) is
0.45(0.51)pb.  All masses are in GeV and cross-sections in
picobarns.

\vskip 12pt
\begin{flushleft}
{\bf Table-II :}
\end{flushleft}

The comparison of the response of the signal to two sets of cuts
{\bf A} and {\bf B}[18] at $\sqrt{s}$=350 GeV where {\rm Cut} \,
{\rm {\bf A}} $\equiv$ 65$<E_{\gamma}<$150 GeV,
40$^\circ<\theta{\gamma}<$140$^\circ$; {\rm Cut} \,{\rm {\bf B}}
$\equiv$ 1$<E_{\gamma}<$150 GeV,
10$^\circ<\theta{\gamma}<$170$^\circ$, $p_{T_{\gamma}}>$10 GeV.
Other fixed values of the SUSY parameters used are
$(\mu,\GLUM,\tanbeta)$=($-$500 GeV, 400 GeV, 2) and
$\MSELL=\MSELR$.  The SM background with Cut {\bf A}({\bf B}) is
0.07671(1.04496)pb.  All masses are in GeV and cross-sections in
picobarns.

\vskip 12pt
\begin{flushleft}
{\bf Table-III :}
\end{flushleft}

The comparison of the response of the signal to two sets of cuts {\bf
A} and {\bf B}[18] at $\sqrt{s}$=500 GeV where {\rm Cut} \, {\rm {\bf
A}} $\equiv$ 95$<E_{\gamma}<$225 GeV,
40$^\circ<\theta{\gamma}<$140$^\circ$; {\rm Cut} \,{\rm {\bf B}}
$\equiv$ 1$<E_{\gamma}<$225 GeV,
10$^\circ<\theta{\gamma}<$170$^\circ$, $p_{T_{\gamma}}>$10 GeV .
Other fixed values of the SUSY parameters used are
$(\mu,\GLUM,\tanbeta)$=($-$500 GeV, 450 GeV, 2) and $\MSELL=\MSELR$.
The SM background with Cut {\bf A}({\bf B}) is 0.07227(1.48773)pb.
All masses are in GeV and cross-sections in picobarns.

\vskip 12pt
\begin{flushleft}
{\bf Table-IV :}
\end{flushleft}

Total signal cross section at $\sqrt{s}=$190 and 350 GeV in
$N=$1 SUGRA model using Cut {\bf A} of Tables I and II. The SUSY
parameters consistent with the VLSP scenario are chosen from
[7]. The underlined entries correspond to the representative
choice $\mu=-\GLUM$ leading to radiative breaking of $SU(2)
\otimes U(1)$ symmetry.  All masses are in GeV and
cross-sections in picobarns.

\vskip 12pt
\begin{flushleft}
{\bf Table-V :}
\end{flushleft}

Effect of soft photon radiative correction at LEP-2 energies.
For each $\MSNU$ the cross sections with and without radiative
corrections are presented using the cut {\bf A} of Table-I. The
choice of SUSY parameters is ($\mu,\GLUM,\tanbeta$)=($-$300 GeV,
200 GeV, 10).  All masses are in GeV and cross-sections in
picobarns.

\newpage

\def \MSNU{m_{\tilde{\nu}}}
\def \GLUM{m_{\tilde{g}}}
\def \tanbeta{{\rm tan}\beta}

\section*{Figure Captions}
\renewcommand{\labelenumi}{Fig. \arabic{enumi}}
\begin{enumerate}

\vspace{10mm}
\item    
\noindent
The region in the  $\MSNU-\GLUM$  plane compatible with the VLSP
scenario with $\tanbeta$ =10  for {\bf (A)} $\mu=-$250 and {\bf
(B)} 250 GeV (see section 2 for further explanations).

\item   
\noindent
photon at $\sqrt{s}=$190 GeV, with $E_{\gamma}>5$ GeV and
$5^{\circ}<\theta_{\gamma}<175^{\circ}$.  {\bf (B)} Angular
distribution of the photon at $\sqrt{s}=$190 GeV with
5$<E_{\gamma}<$60 GeV and 5$^\circ < \theta_{\gamma} <$175
$^\circ$.  The convention for different lines and SUSY
parameters chosen are explained in section 2.

\item   
\noindent
The total  cross section (SM+VLSP) as a function of $\MSNU$ at
$\sqrt{s}=$190 GeV. The band within the solid lines correspond
to the SM+$\tilde{\nu} \tilde{\bar{\nu}}+\widetilde{N}_1
\widetilde{N}_1$ cross section and is obtained by varying
$\GLUM$, $\mu$ and $\tanbeta$ over the LEP-1 allowed region.
The band within the dashed lines is obtained by taking
additional contributions from $\widetilde{N}_1 \widetilde{N}_2$
pairs into account.  The horizontal dotted lines correspond to
the SM background and its fluctuations (see section 2).

\item   
\noindent
Contour plots in the $(\MSNU-\GLUM)$ plane at $\sqrt{s}=$190 GeV
indicating the regions where $\geq$2$\sigma$(dotted),
$\geq$3$\sigma$(dashed) and $\geq$4$\sigma$(solid) signals may
be obtained for $\tanbeta=$ 2,10 and 30 and $-500 \leq \mu
\leq$500.  The shaded region corresponds to $\MSNU < \MCH1 <
\MSNU + 5$ GeV, where the chargino decay is likely to be
difficult to detect.

\item   
\noindent
{\bf (A)} Energy distribution of the photon at $\sqrt{s}=$350
GeV, with $E_{\gamma}>$5 GeV and
$5^{\circ}<\theta_{\gamma}<175^{\circ}$.  {\bf (B)} Angular
distribution for the photon at $\sqrt{s}=$350 GeV with
$15<E_{\gamma}<150$ and $5^\circ < \theta_{\gamma} <175 ^\circ$.
The convention for different lines and SUSY parameters chosen
are explained in section 3.

\item   
\noindent
The total  cross section (SM+VLSP) as a function of $\MSNU$ at
$\sqrt{s}=$350 GeV. The conventions for different bands and
lines are explained in Fig.[3] and in section 3.

\newpage
\item   
\noindent
Contour plots in the $(\MSNU-\GLUM)$ plane at $\sqrt{s}=$350 GeV
indicating the regions where $\geq$3$\sigma$(dotted),
$\geq$4$\sigma$(dashed) and $\geq$5$\sigma$(solid) signals may
be obtained for $\tanbeta=$ 2,10 and 30 and $-$500$\leq \mu
\leq$500.

\item   
\noindent
{\bf (A)} Energy distribution of the photon at $\sqrt{s}=$500
GeV, with $E_{\gamma}>5$ GeV and
$5^{\circ}<\theta_{\gamma}<175^{\circ}$.  {\bf (B)} Angular
distribution for the photon at $\sqrt{s}=$500 GeV with
$25<E_{\gamma}<225$ and $5^\circ < \theta_{\gamma} <175^\circ$.
The convention for different lines and SUSY parameters chosen
are explained in the text.

\item   
\noindent
The total cross section (SM+VLSP)  as a function of $\MSNU$ at
$\sqrt{s}=$500 GeV. The conventions are the same as in
Fig.[6].

\item   
\noindent
Contour plots in the $(\MSNU-\GLUM)$ plane at $\sqrt{s}=$500 GeV
following the conventions of Fig.[7].

\item   
\noindent
Feynman diagrams for the process $e^+ \, e^- \longrightarrow \nu
\bar{\nu} \, + \, \gamma $.  In this and in the subsequent figures,
the index $l$ stands for $e,\mu,\tau$. Arrows indicate the flow of
physical momenta and not of fermion numbers. 

\item   
\noindent
Feynman diagrams for the process $e^+ \, e^- \longrightarrow
\tilde{\nu} \tilde{\bar{\nu}} \, + \, \gamma $.

\item   
\noindent
Feynman diagrams for the process $e^+ \, e^- \longrightarrow
\widetilde{N}_i \widetilde{N}_j \, + \, \gamma $.

\end{enumerate}

\newpage

\def \MSNU{m_{\tilde{\nu}}}
\def \GLUM{m_{\tilde{g}}}
\def \tanbeta{{\rm tan}\beta}
\def \MSQUARK{m_{\tilde{q}}}

\vskip 5pt
\begin{center}
{\bf Table-I}
\end{center}
\begin{center}
\begin{tabular}{||c|c|c|c|c|c|c||}
\hline
\hline
$m_{\tilde{\nu}}$ & Cut & $\sigma_{\tilde{\nu} \tilde{\nu}}$ &
$\sigma_{\tilde{N}_1 \tilde{N}_1}$ & $\sigma_{\tilde{N}_1 
\tilde{N}_2}$ & Total & $\sigma = \frac{S}{\sqrt{B}}$\\
\hline
\hline
45 & {\bf A} & .103 & .027 & .018 & .148 & 4.9\\
   & {\bf B} & .117 & .031 & .022 & .170 & 5.3\\
\hline
55 & {\bf A} & .077 & .025 & .016 & .118 & 3.9\\
   & {\bf B} & .087 & .029 & .019 & .135 & 4.2\\
\hline
65 & {\bf A} & .048 & .023 & .015 & .086 & 2.8\\
   & {\bf B} & .053 & .026 & .017 & .096 & 3.0\\
\hline
\hline
\end{tabular}
\end{center}

\vskip 20pt
\begin{center}
{\bf Table-II}
\end{center}
\begin{center}
\begin{tabular}{||c|c|c|c|c|c|c|c||}
\hline
\hline
$m_{\tilde{\nu}}$ & Cut & $\sigma_{\tilde{\nu} \tilde{\nu}}$ &
$\sigma_{\tilde{N}_1 \tilde{N}_1}$ & $\sigma_{\tilde{N}_1
\tilde{N}_2}$ & $\sigma_{\tilde{N}_2 \tilde{N}_2}$ &Total &
$\sigma = \frac{S}{\sqrt{B}}$\\
\hline
\hline
110 & {\bf A} & .00156 & .00212 &  disallowed$^!$ &
disallowed$^!$ &
.00368 & 2.3 \\ 
    & {\bf B} & .02037 & .01803 &  &  & .03840 & 6.7 \\
\hline
125 & {\bf A} & .00041 & .00182 & disallowed$^!$ &
disallowed$^!$ & .00223 & 1.4\\ & {\bf B} & .01194 & .01600 &  &
& .02794 & 4.8\\
\hline
130 & {\bf A} & .00018 & .00173 & .00143 & negligible & .00334 & 2.2\\
    & {\bf B} & .00948 & .01539 & .01397 & .00101 & .03985 & 6.9\\
\hline
135 & {\bf A} & .00004 & .00165 & .00134 & negligible & .00303 & 1.9\\
    & {\bf B} & .00721 & .01476 & .01324 & .00096 & .03617 & 6.3\\
\hline
$^\ast$150 & {\bf A} & negligible & .00113 & .00074 & negligible
& .00187 &1.2 \\ & {\bf B} & .00173 & .01162 & .01060 & .00013 &
.02408 & 4.1 \\
\hline
\hline
\end{tabular}
\end{center}
\vskip 3pt $^\ast$ In this case $\GLUM$=450 GeV, since $\GLUM$ =
400 GeV is not allowed in the VLSP scenario.\\ $^!$
$\widetilde{N}_2$ cannot be a VLSP for this choice of SUSY
parameters.\\

\newpage
\begin{center}
{\bf Table-III}
\end{center}
\begin{center}
\begin{tabular}{||c|c|c|c|c|c|c|c||}
\hline
\hline
$m_{\tilde{\nu}}$ & Cut & $\sigma_{\tilde{\nu} \tilde{\nu}}$ &
$\sigma_{\tilde{N}_1 \tilde{N}_1}$ & $\sigma_{\tilde{N}_1
\tilde{N}_2}$ & $\sigma_{\tilde{N}_2 \tilde{N}_2}$ &Total &
$\sigma = \frac{S}{\sqrt{B}}$\\
\hline
\hline
150 & {\bf A} & .00138 & .00156 & .00130 & .00012 & .00436 & 2.9 \\
    & {\bf B} & .02207 & .01417 & .01114 & .00344 & .05082 & 7.4 \\
\hline
155 & {\bf A} & .00113 & .00150 & .00124 & .00012 & .00399 & 2.6 \\
    & {\bf B} & .02030 & .01378 & .01073 & .00333 & .04814 & 7.0 \\
\hline
$^ \dagger$ 160 & {\bf A} & .00078 & .00129 & disallowed$^!$ &
disallowed$^!$ & .00207 & 1.4 \\ 
    & {\bf B} & .01582 & .01254 &  &  & .02836 & 4.1 \\
\hline
$^ \dagger$ 165 & {\bf A} & .00060 & .00124 & .00106 & .00004 &
.00294 & 1.9 \\ & {\bf B} & .01430 & .01218 & .01013 & .00231 &
.03892 & 5.7 \\
\hline
$^\dagger$ 170 & {\bf A} & .00044 & .00120 & .00101 & .00003 &
.00268 & 1.8 \\ & {\bf B} & .01287 & .01185 & .00975 & .00223 &
.03670 & 5.3 \\
\hline
$^\ddagger$ 180 & {\bf A} & .00017 & .00098 & .00083 &
negligible & .00198 & 1.3 \\ & {\bf B} & .00852 & .01048 &
.00918 & .00136 & .02954 & 4.3 \\
\hline
\end{tabular}
\end{center}
$^\dagger \GLUM$=500 GeV \\
$^!$ $\widetilde{N}_2$ cannot be a VLSP for this choice of SUSY
parameters.  \\
$^\ddagger \GLUM$=550 GeV.\\

\newpage
\begin{center}
{\bf Table-IV}
\end{center}
\begin{center}
\begin{tabular}{||c|c|c|c|c|c|c|c|c||}
\hline
\hline
$\sqrt{s}$ & $\tanbeta$ & $m_0$ & $M_2$ & $\mu$ & $\GLUM(\MSQUARK)$ &
$\MSNU$ & $\sigma_{SUSY}$ & $\sigma=\frac{S}{\sqrt{B}}$ \\ 
\hline
\hline 
& & 20 & 75 & -55 & 225(225) & 47.5 & .141 & 4.7\\
& & 40 & 70 & -65 & 210(213) & 53.8 & .108 & 3.6\\
190 & 2 & 30 & 70 & 320 & 210(211) & 46.8 & .219 & 7.3\\
& & 40 & 70 & 400 & 210(213) & 53.8 & .170 & 5.6\\
& & 50 & 70 & 700 & 210(215) & 61.6 & .124 & 4.1\\
& & 50 & 75 & 800 & 225(230) & 66.0 & .099 & 3.3 \\ 
\hline
& & 20 & 75 & -55 & 225(225) & 47.5 & .03478 & 22.3\\
& & 40 & 80 & -75 & 240(242) & 63.7 & .02338 & 15.0\\
& & 75 & 70 & -125 & 210(222) & 83.2 & .01838 & 11.8\\
350& 2 & 30 & 70 & 320 & 210(211) & 46.8 & .05099 & 32.8\\
& & 40 & 70 & 500 & 210(213) & 53.8 & .04164 & 26.8\\
& & 50 & 70 & 1000 & 210(215) & 61.6 & .03467 & 22.3\\
& & 55 & 80 & 1000 & 240(245) & 74.0 & .02494 & 16.0\\
\hline
& & 60 & 70 & -160 & 210(218) & 57.6 & .121 & 4.1\\
& & 60 & 70 & \underbar{-210} & \underbar{210(218)} & 57.6 &
\underbar{.118} & 3.9\\ 
& & 40 & 80 & -120 & 240(242) & 49.7 & .144 & 4.8\\
190 & 10 & 40 & 90 & -120 & 270(272) & 61.5 & .094 & 3.1\\
& & 20 & 90 & 120 & 270(270) & 50.9 & .155 & 5.1\\
& & 60 & 70 & 280 & 210(218) & 57.6 & .128 & 4.2\\
& & 60 & 70 & 500 & 210(218) & 57.6 & .121 & 4.0\\
\hline
& & 60 & 70 & \underbar{-210} & \underbar{210(218)} & 57.6 &
\underbar{.03388} & 21.8\\ 
& & 60 & 80 & -60 & 240(246) & 66.9 & .02648 & 17.0\\
& & 60 & 80 & \underbar{-240} & \underbar{240(246)} & 66.9 &
\underbar{.02605} & 16.7\\ 
& & 60 & 90 & -200 & 270(276) & 76.1 & .02014 & 12.9\\
350 & 10 & 75 & 100 & \underbar{-300} & \underbar{300(308)} & 96.3 &
\underbar{.01169} & 7.5\\ 
& & 40 & 90 & -120 & 270(272) & 61.5 & .02789 & 17.9\\
& & 40 & 90 & 160 & 270(272) & 61.5 & .02990 & 19.2\\
& & 60 & 100 & 280 & 300(305) & 85.2 & .01292 & 8.3\\
& & 70 & 85 & 500 & 255(263) & 80.1 & .01960 & 12.6\\
\hline
\hline
\end{tabular}
\end{center}

\newpage
\vskip 5pt
\begin{center}
{\bf Table-V}
\end{center}
\begin{center}
\begin{tabular}{||c|c|c|c|c|c||}
\hline
\hline
$m_{\tilde{\nu}}$ & $\sigma_{\tilde{\nu} \tilde{\nu}}$ &
$\sigma_{\tilde{N}_1 \tilde{N}_1}$ & $\sigma_{\tilde{N}_1 \tilde{N}_2}$ &
Total & $\sigma = \frac{S}{\sqrt{B}}$\\
\hline
\hline
45 & .103 & .027 & .018 & .148 & 4.9\\
   & .089 & .020 & .014 & .123 & 4.7\\
\hline
55 & .077 & .025 & .016 & .118 & 3.9\\
   & .057 & .019 & .013 & .089 & 3.4\\
\hline
65 & .048 & .023 & .015 & .086 & 2.8\\
   & .034 & .017 & .011 & .062 & 2.4\\
\hline
\hline
\end{tabular}
\end{center}

\newpage
\thispagestyle{empty}
\begin{figure}[htb]
\epsffile[50 350 0 780]{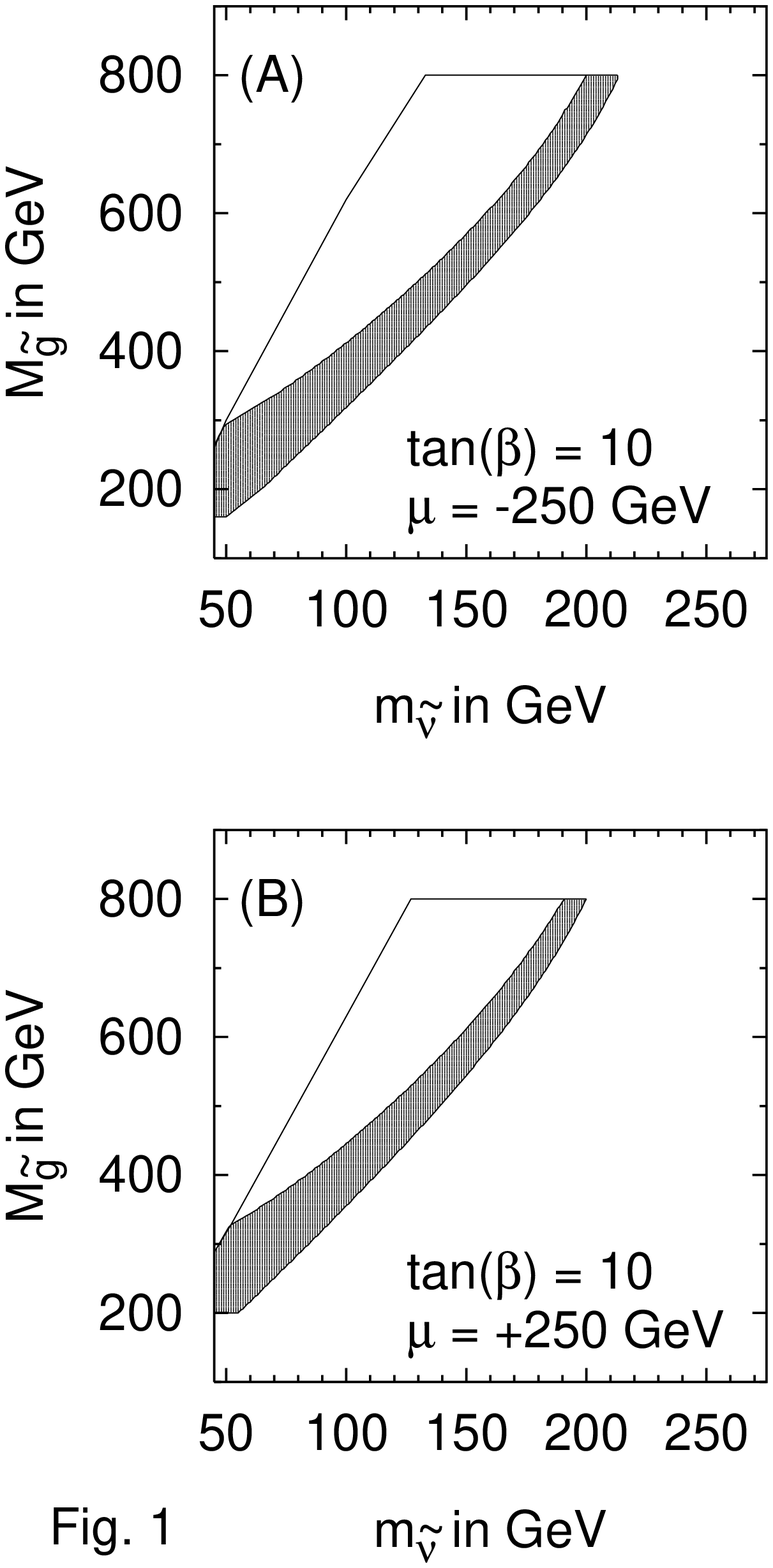}
\end{figure}

\newpage
\thispagestyle{empty}
\begin{figure}[htb]
\epsffile[50 350 0 780]{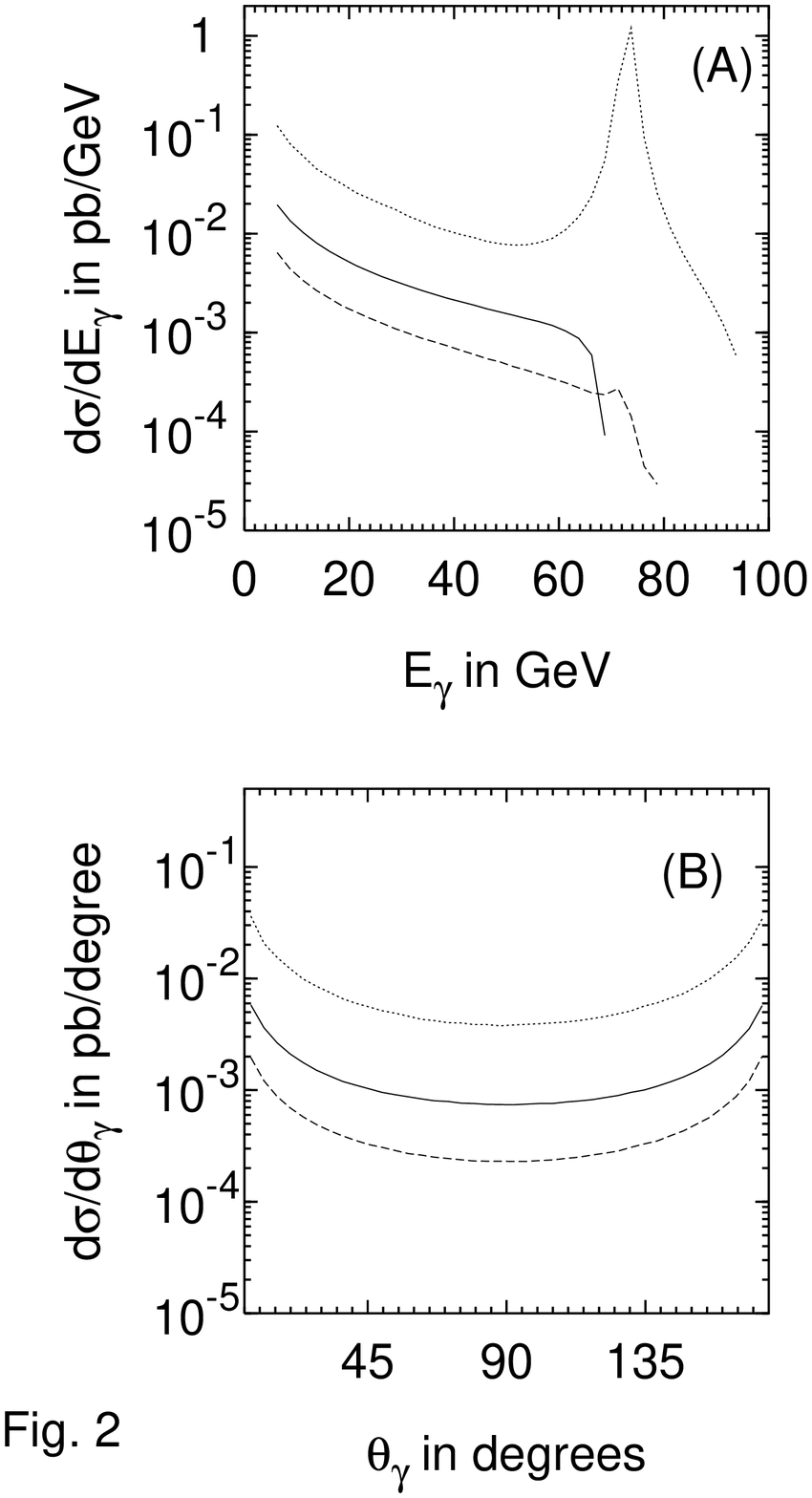}
\end{figure}

\newpage
\thispagestyle{empty}
\begin{figure}[htb]
\epsffile[50 350 0 780]{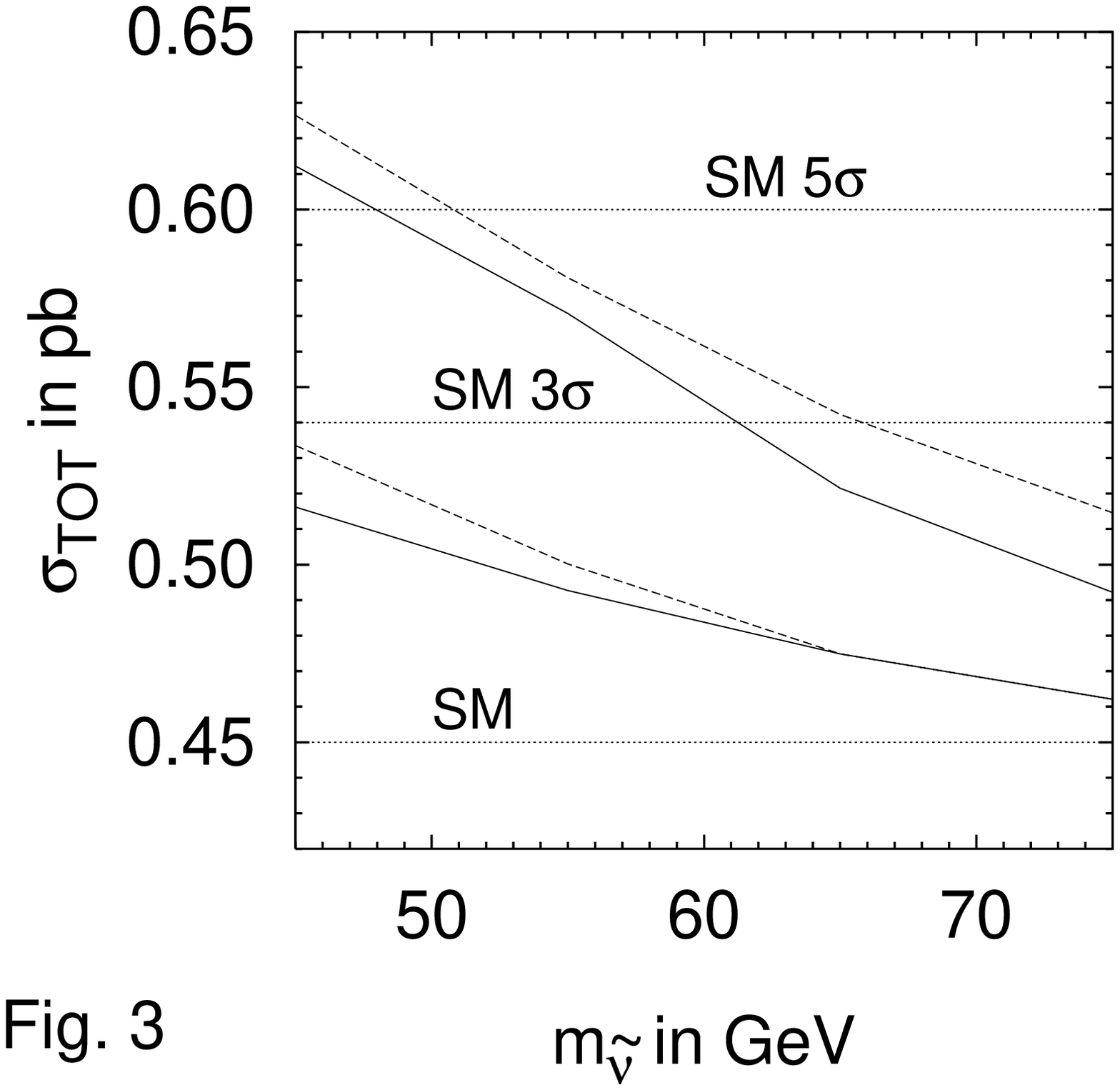}
\end{figure}

\newpage
\thispagestyle{empty}
\begin{figure}[htb]
\epsffile[50 350 0 780]{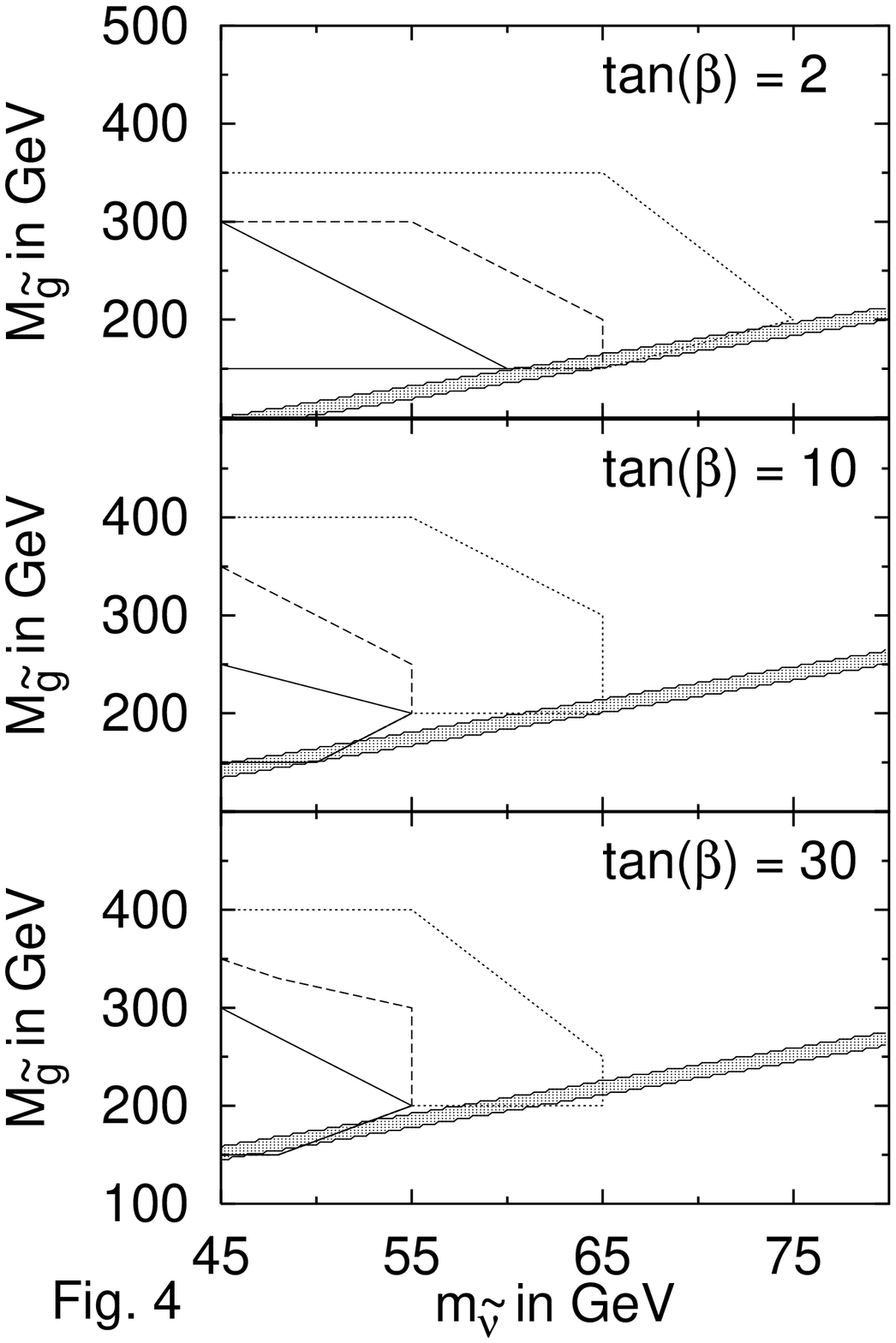}
\end{figure}

\newpage
\thispagestyle{empty}
\begin{figure}[htb]
\epsffile[50 350 0 780]{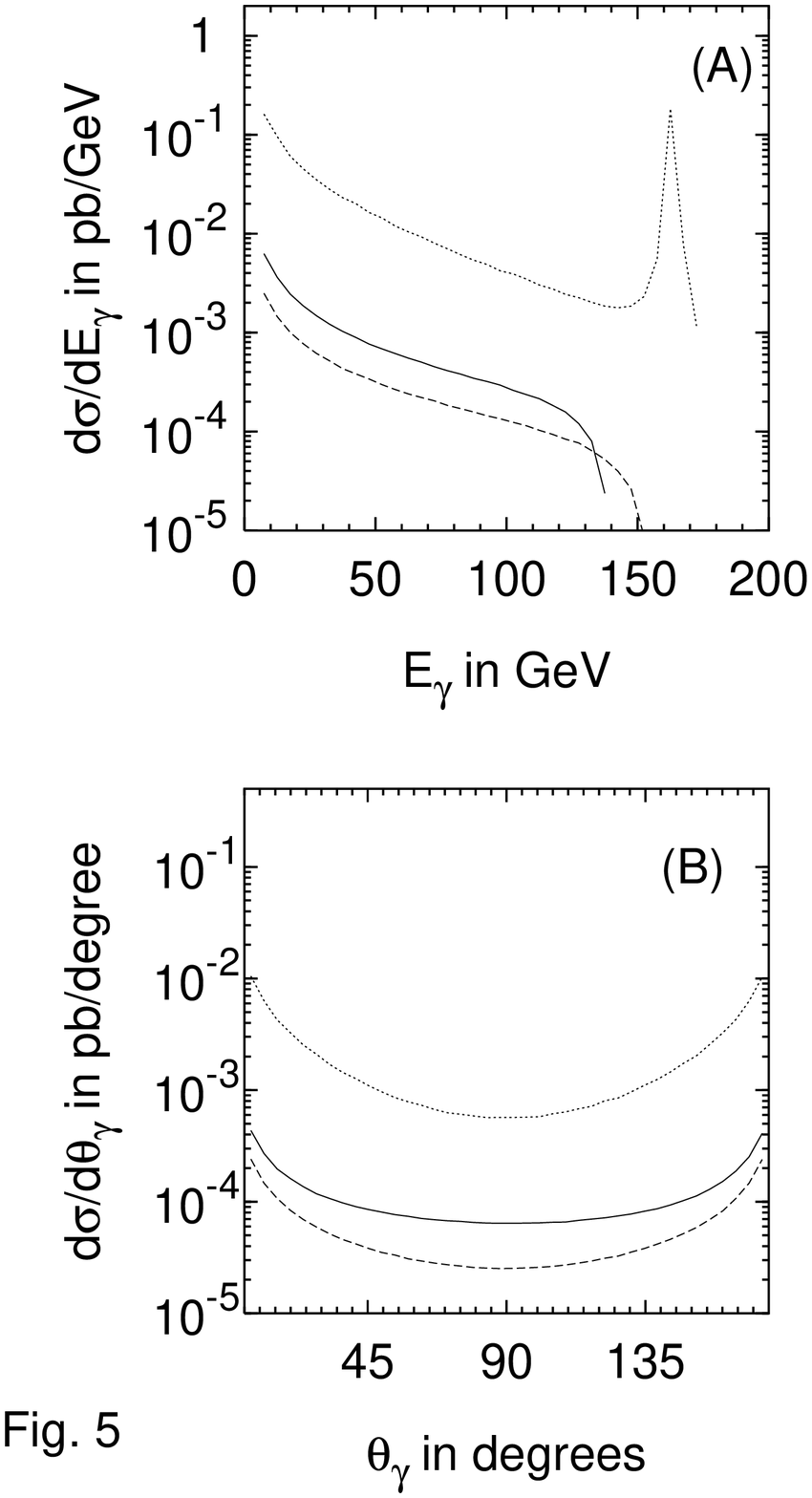}
\end{figure}

\newpage
\thispagestyle{empty}
\begin{figure}[htb]
\epsffile[50 350 0 780]{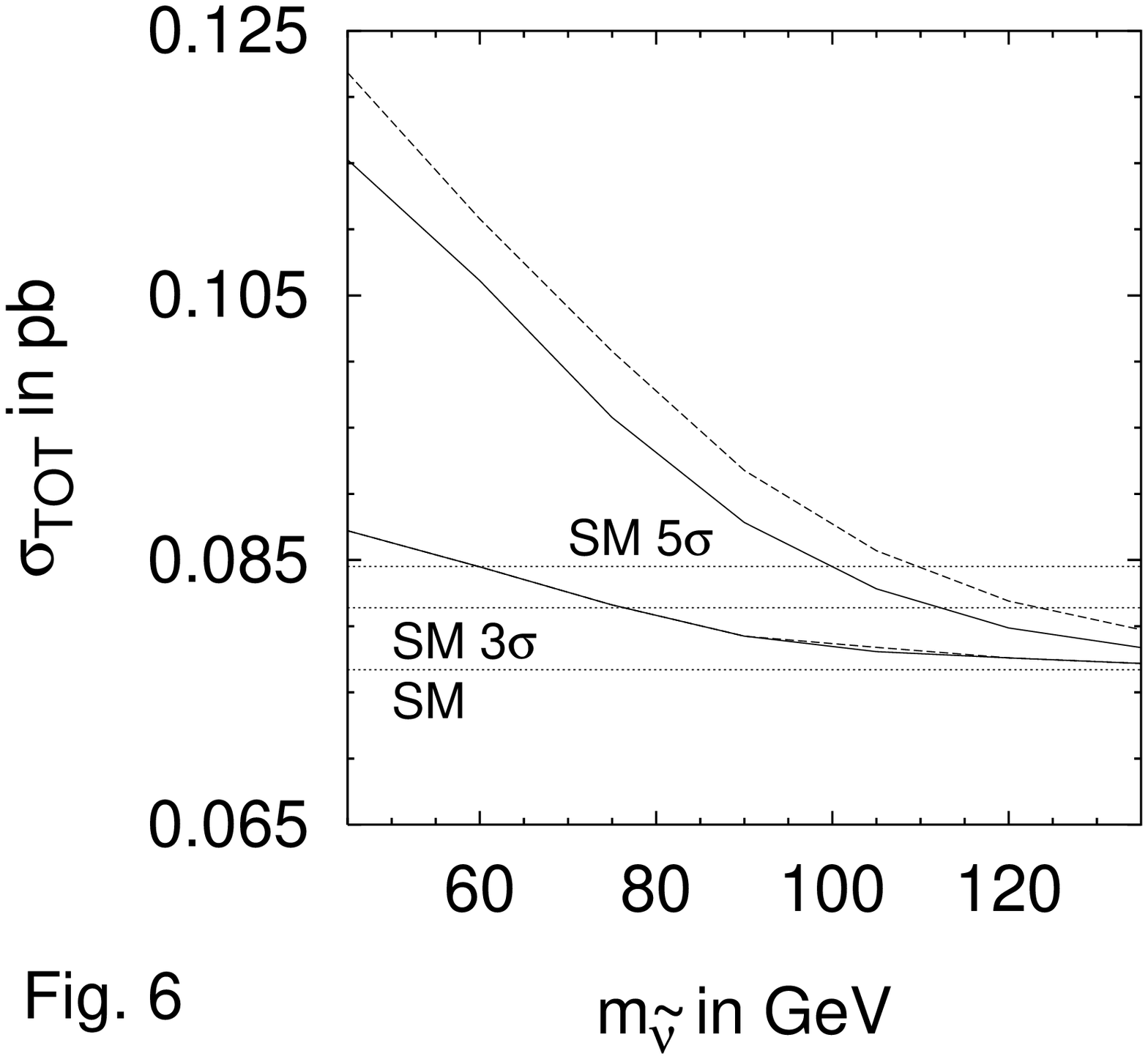}
\end{figure}

\newpage
\thispagestyle{empty}
\begin{figure}[htb]
\epsffile[50 350 0 780]{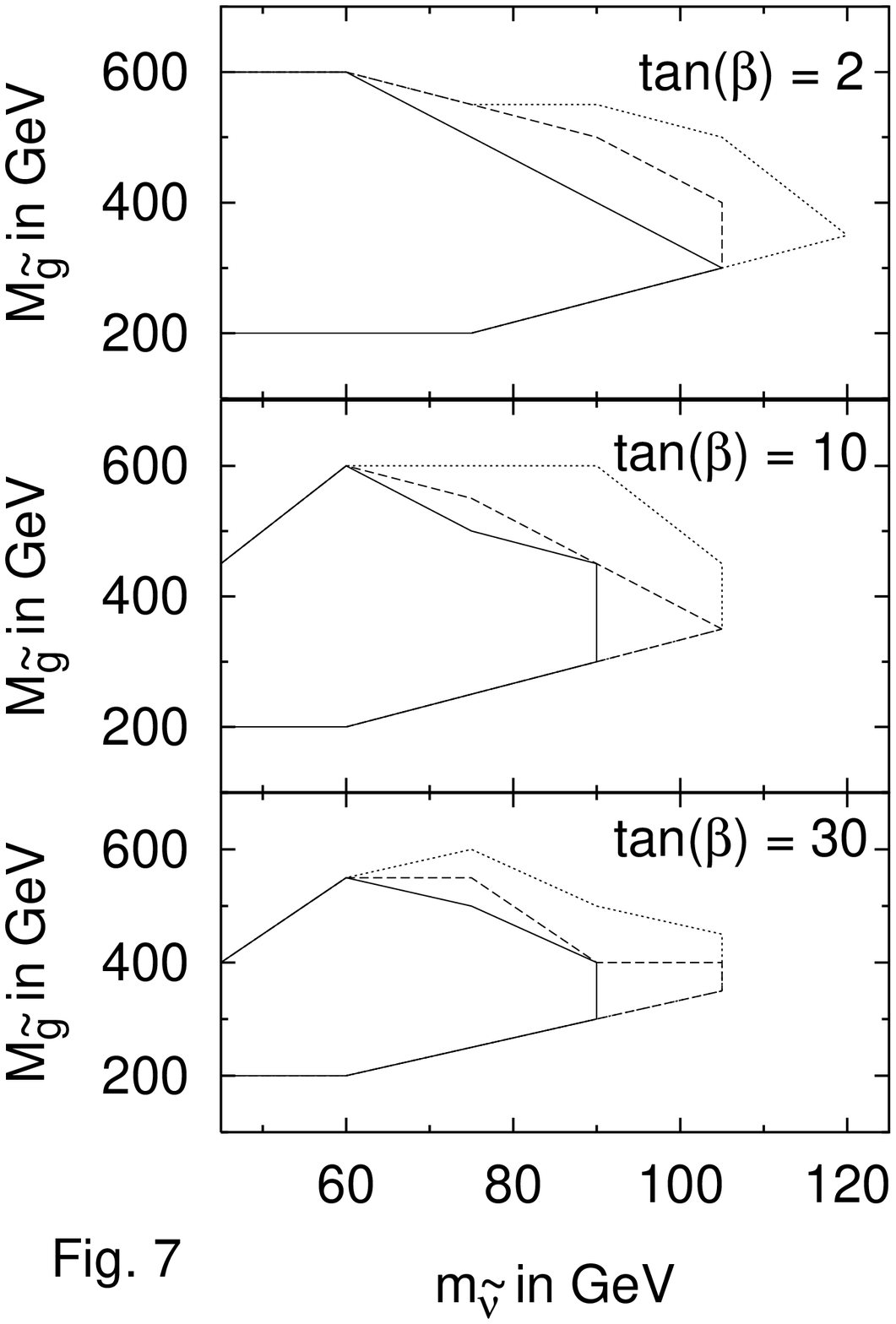}
\end{figure}

\newpage
\thispagestyle{empty}
\begin{figure}[htb]
\epsffile[50 350 0 780]{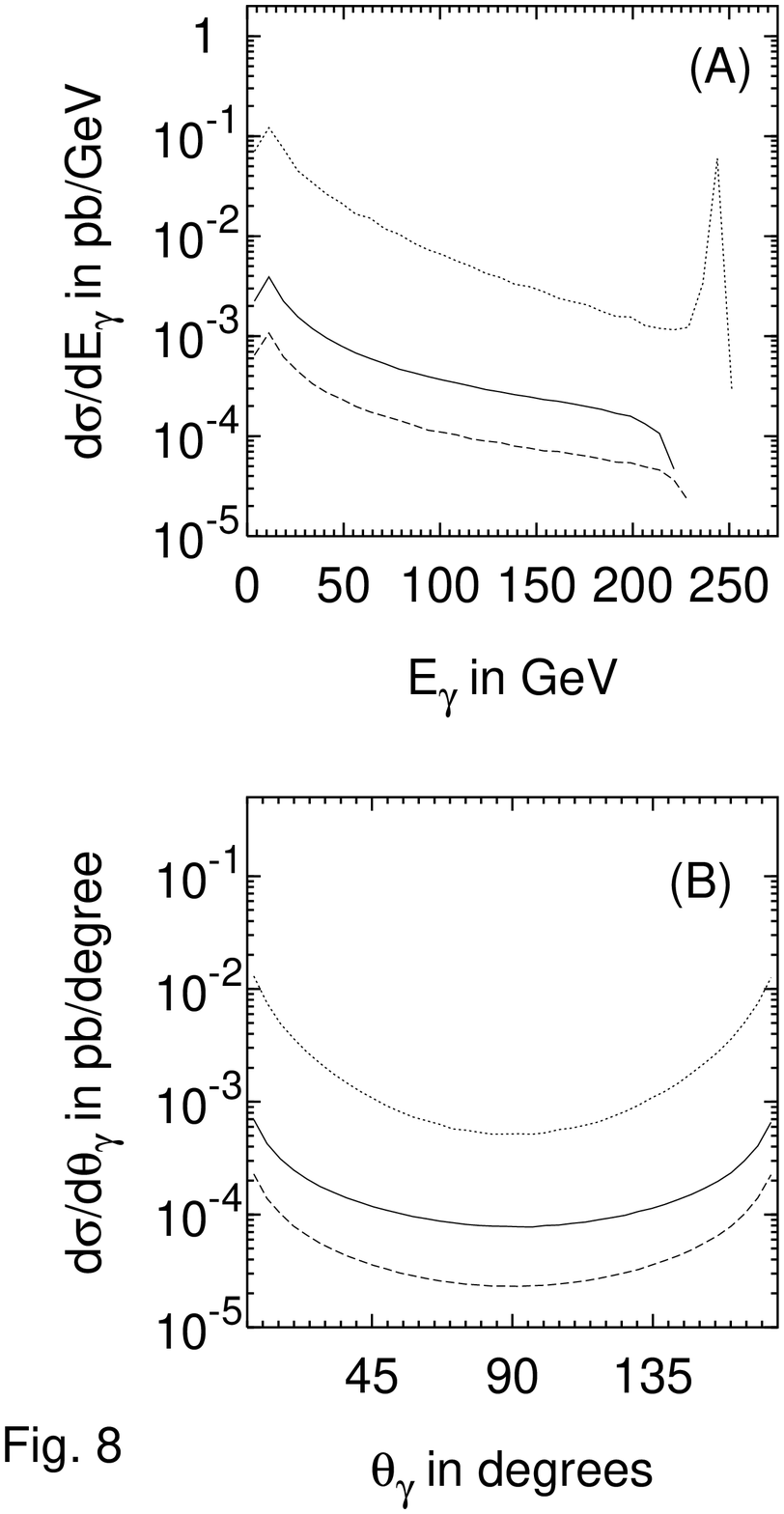}
\end{figure}

\newpage
\thispagestyle{empty}
\begin{figure}[htb]
\epsffile[50 350 0 780]{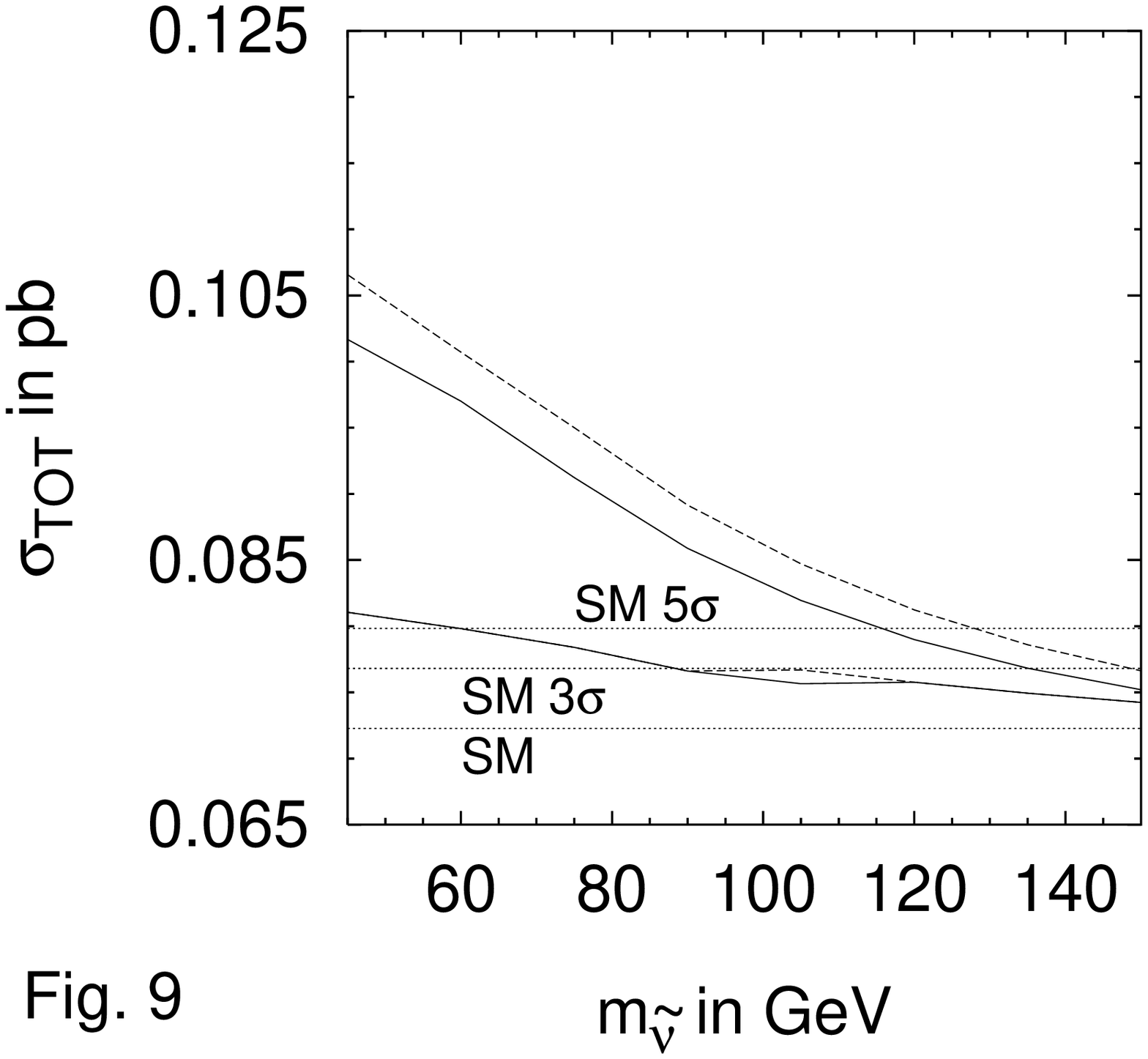}
\end{figure}

\newpage
\thispagestyle{empty}
\begin{figure}[htb]
\epsffile[50 350 0 780]{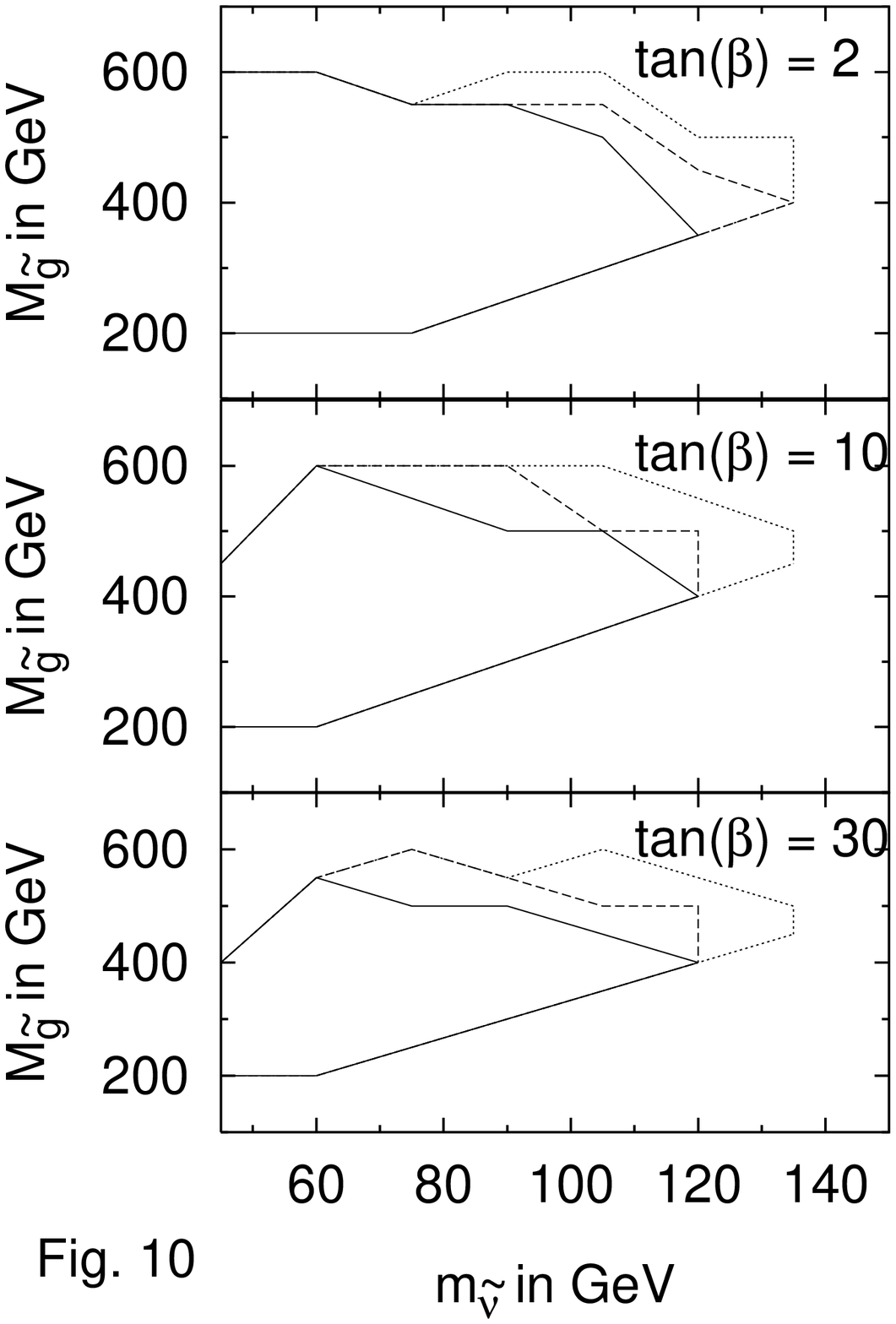}
\end{figure}

\newpage
\thispagestyle{empty}
\begin{figure}[htb]
\epsffile[50 350 0 780]{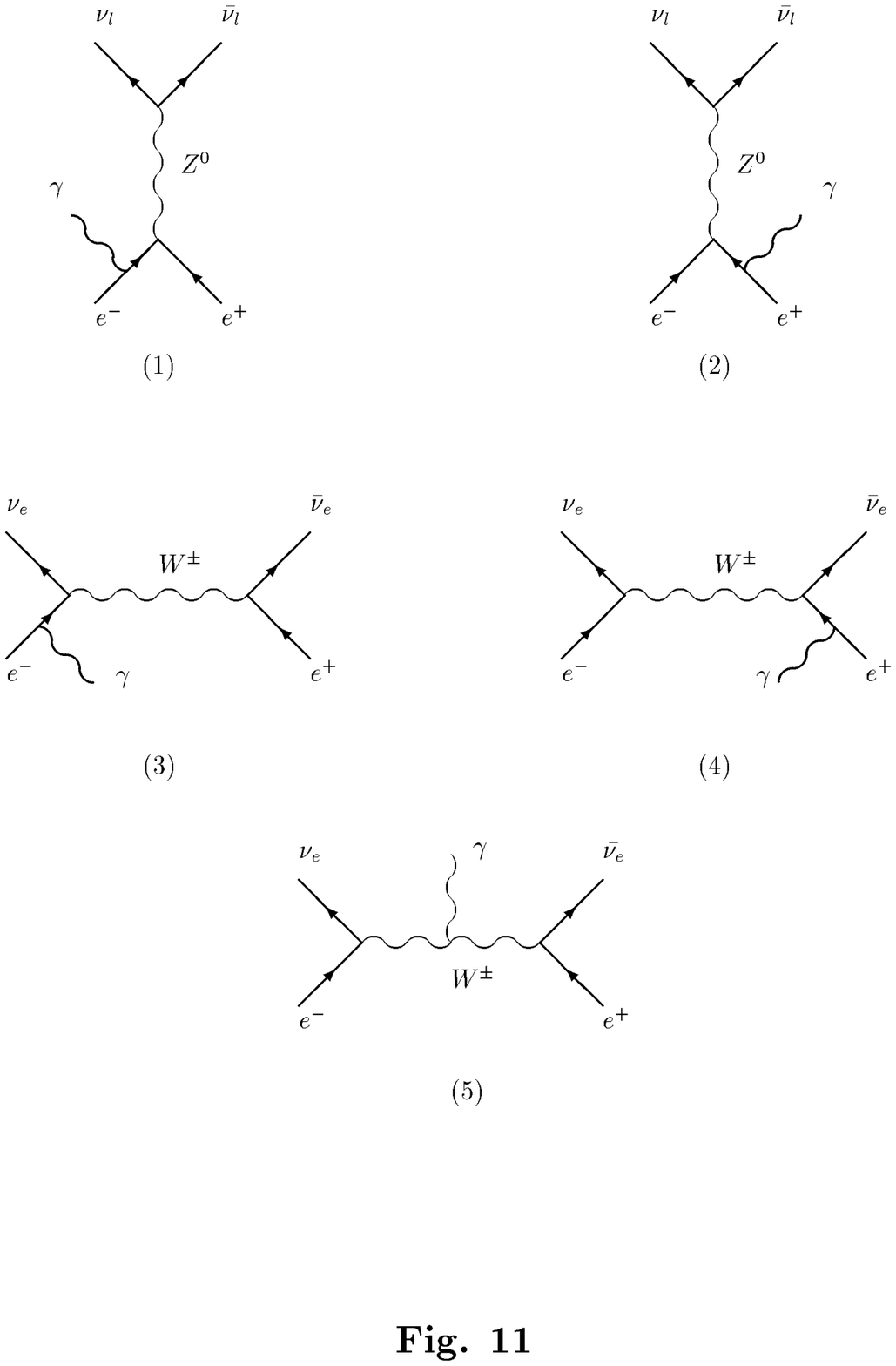}
\end{figure}

\newpage
\thispagestyle{empty}
\begin{figure}[htb]
\epsffile[50 350 0 780]{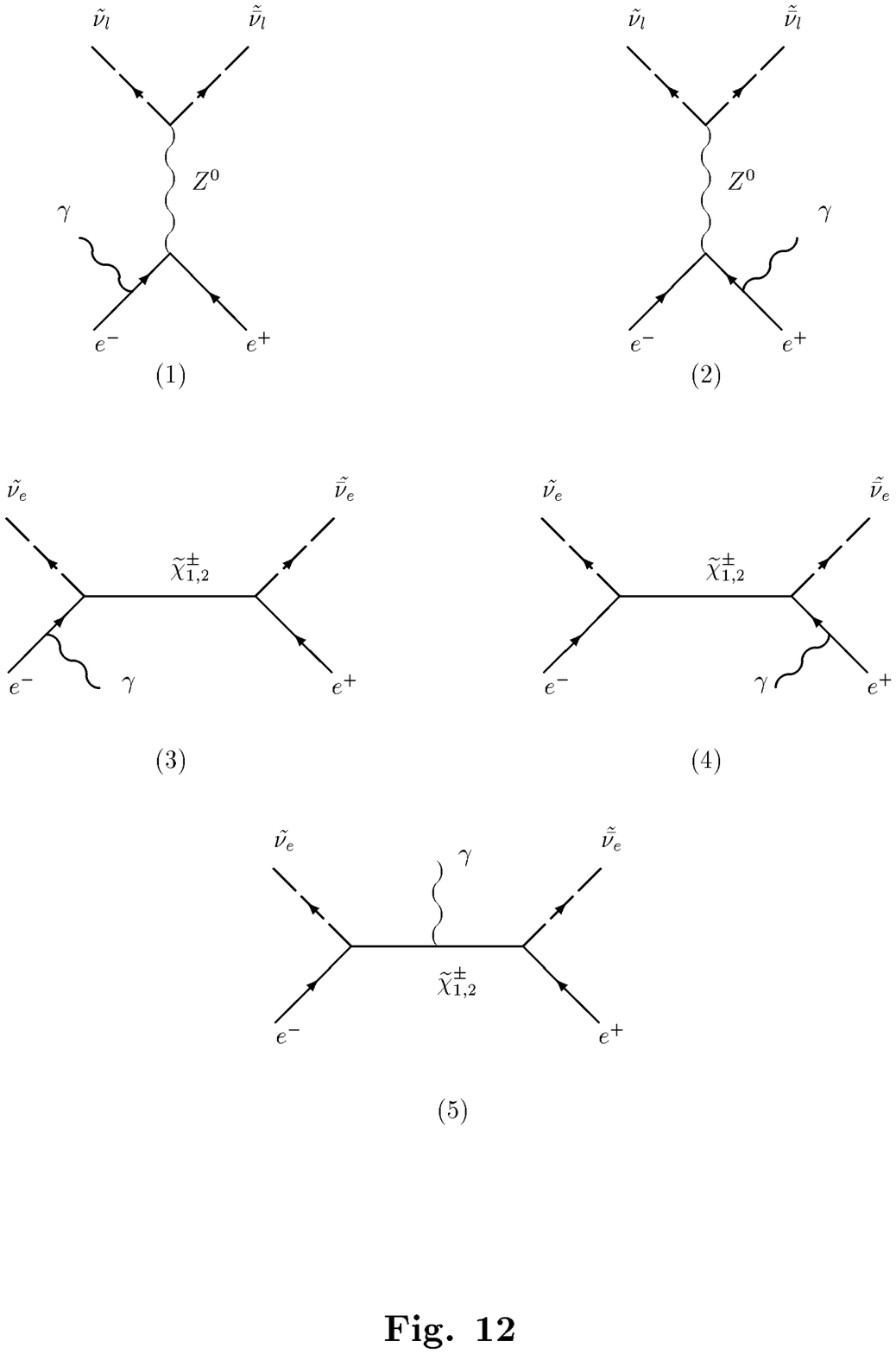}
\end{figure}

\newpage
\thispagestyle{empty}
\begin{figure}[htb]
\epsffile[50 350 0 780]{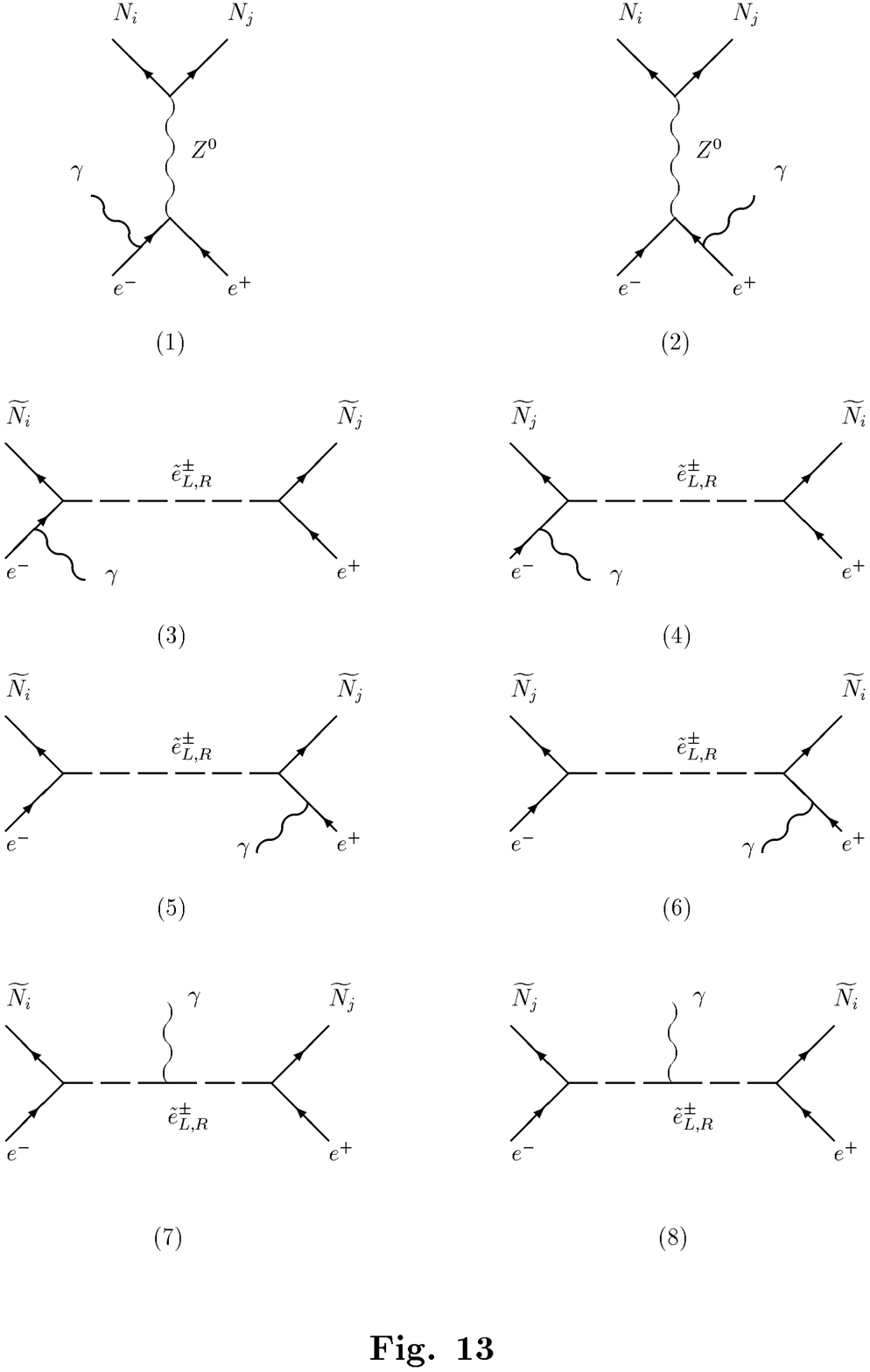}
\end{figure}

\end{document}